%% file: anonymous-submission-latex-2026.tex
\documentclass[letterpaper]{article} 
\usepackage{aaai2026}  
\usepackage{times}  
\usepackage{helvet}  
\usepackage{courier}  
\usepackage[hyphens]{url}  
\usepackage{graphicx} 
\usepackage{booktabs}
\usepackage{natbib}  
\usepackage{caption} 
\usepackage{algorithm}
\usepackage{algorithmic}

\usepackage{xcolor} 
\newcommand{\answerYes}[1]{\textcolor{blue}{#1}} 
 
\newcommand{\answerNA}[1]{\textcolor{gray}{#1}}

\usepackage{newfloat}
\usepackage{listings}
\DeclareCaptionStyle{ruled}{labelfont=normalfont,labelsep=colon,strut=off} 
\floatstyle{ruled}
\newfloat{listing}{tb}{lst}{}
\floatname{listing}{Listing}
\usepackage{enumitem}
\usepackage{multirow}
\usepackage{amsmath}
\usepackage{amssymb}
\usepackage{url}

\title{Framing War Across Languages: \\Power, Agency, and Sentiment in Wikipedia's Multilingual War Narratives}
\author{
    Jiarui Xia,
    Diego Gomez-Zara
}

\affiliations{
    University of Notre Dame\\
    Notre Dame, USA\\
    jxia3@nd.edu, dgomezza@nd.edu
}

\begin{document}

\maketitle

\begin{abstract}
\input{00_Abstract}
\end{abstract}

\input{01_Introduction}
\input{02_RelatedWork}
\input{03_ResearchQuestions}
\input{04_Data}
\input{05_Methodology}
\input{06_Results}
\input{07_Discussion}


\bibliography{aaai2026}
\newpage
\input{08_EthicsChecklist}

\input{09_Appendix}

\end{document}

%% file: 00_Abstract.tex
While Wikipedia promotes a neutral point of view on historical conflicts, its language editions are written by editors from distinct linguistic and cultural communities. In this study, we analyze 158 wars since 1900 to examine how the descriptions of combatants vary across 20 Wikipedia language editions. Using connotation frames---which assess power, agency, and sentiment toward an entity---we examine how each language portrays the parties involved in the conflict. We find systematic differences when language editions describe wars involving their own communities, although the direction of these asymmetries varies across languages. However, when language editions describe conflicts that do not involve their own linguistic communities, their narrative structures exhibit high cross-linguistic similarity. These findings show how linguistic communities influence war narratives on Wikipedia, revealing that shared historical accounts remain shaped by the perspectives of the language communities that produce them.


%% file: 01_Introduction.tex
\section{Introduction}
Wikipedia is one of the world's most influential sources of information about history and conflict, aspiring to present balanced accounts across hundreds of language editions through its Neutral Point of View (NPOV) policy\footnote{\scriptsize{\url{https://en.wikipedia.org/wiki/Wikipedia:Neutral_point_of_view}}}. Yet, Wikipedia is not a single unified system but a collection of hundreds of language editions, each maintained by independent editor communities embedded in distinct cultural, linguistic, and geopolitical contexts \cite{hecht2010tower, johnson2022considerations}. One prominent pattern is a self-focus bias, whereby language communities tend to prioritize topics, individuals, and events that are more closely related to their own cultural and regional contexts \cite{miquel2011self,oeberst2016individual}. As a result, language editions---even when describing the same events---often differ in perspective, framing, and information selection.

These differences become particularly salient in the context of war. The same conflict may be narrated in markedly different ways depending on national context, including how responsibility is assigned and actors are portrayed \cite{entman1993framing,halbwachs2024collective,wertsch2008collective}. Prior research has shown that different language editions vary in what they document and emphasize \cite{callahan2011cultural,laufer2015mining,he2018the_tower_of_babel}, and studies of intergroup conflict find ingroup bias in Wikipedia articles \cite{oeberst2020collectively}. However, much of this work relies on coverage-based measures, coarse sentiment analysis, or small-scale qualitative studies \cite{rogers2013wikipedia,ferreira2022construction,gustafsson2020international}, leaving open how combatants are relationally portrayed across languages and how such portrayals depend on a language community's relationship to the conflict. In this study, we ask: \textit{Do Wikipedia language editions systematically differ in how they portray combatant entities in wars, and what explains these differences?}


To address this question, we conduct a large-scale, cross-linguistic analysis of Wikipedia's multilingual war narratives. Focusing on 2,542 battles from 158 post-1900 wars across 20 Wikipedia language editions, we use connotation frames \cite{rashkin2016connotation,sap2017connotation} derived from subject-verb-object relations to measure how combatants are portrayed in terms of power, agency, and sentiment. Building on this framework, we compare how different language editions portray self-aligned and opposing entities in war narratives when the corresponding language communities are directly involved in the conflict. We examine (1) how language editions portray the sides in a war, (2) how power and agency are attributed to aligned versus opposite entities, (3) cultural and editorial factors explaining cross-linguistic differences, and (4) whether narratives converge across languages when communities are not directly involved in the conflict.

Our results show systematic differences in how Wikipedia language editions portray combatant entities. Some portray self-aligned entities as more powerful and agentic, while others portray them as less so. In contrast, when languages describe wars in which they are not directly involved, narrative structures largely converge, indicating that direct involvement is an important source of cross-linguistic divergence. Together, these findings identify conditions under which collaborative knowledge production converges or diverges across linguistic communities.

This work makes two contributions: we extend framing analysis to Wikipedia at scale, demonstrating how narrative framing of war emerges through collaborative knowledge production even under shared neutrality norms; and provide an entity-centered approach for comparing war narratives across languages, moving beyond coverage metrics to analyze how combatants are relationally portrayed.

%% file: 02_RelatedWork.tex
\section{Related Work}

\subsection{War Narratives and Collective Memory}
What societies remember about war is never simply a product of a conflict's scale or historical significance, but rather a reflection of whether that conflict aligns with prevailing political agendas, national identity, and cultural traditions \cite{ashplant2000commemorating, connerton2017seven, halbwachs2024collective}. Wars framed around clear moral distinctions tend to become deeply institutionalized in public consciousness, reproduced across education, commemoration rituals, and political discourse. Conflicts that lack such ideological clarity, or whose outcomes are politically inconvenient, are more susceptible to selective marginalization or erasure \cite{misztal2010collective, connerton2017seven}. This selectivity is actively produced, as states and dominant institutions suppress narratives that challenge national cohesion, while amplifying those that serve legitimation needs \cite{verovvsek2016collective}.

These ideologically shaped memories are constructed and disseminated through media, film, and news outlets that function as primary vehicles of collective remembrance across generations \cite{godfrey2009visual, schudson1992watergate}. Instead of transmitting a fixed historical record, such media reconstruct the past in light of present pressures, selectively elevating certain actors and embedding conflicts within moral frameworks that serve contemporary ideological purposes \cite{baudrillard1994illusion}. This dynamic leads different countries to routinely place divergent emphases on actors, causal chains, and moral evaluations when reporting on the same conflict \cite{baum2015filtering, stawski2025framing}, producing narratives calibrated to legitimize their national identities and political responses \cite{abel2019collective, liu2005past}.


\subsection{Cross-Linguistic Variation in Wikipedia}
Wikipedia comprises over 300 language editions, aiming to represent the same topics across cultural contexts \cite{hecht2010tower, johnson2022considerations}. Prior work has shown systematic cross-linguistic variation in content. For instance, \citet{callahan2011cultural} found that English and Polish Wikipedia articles about famous individuals differed in emphasis on personal details versus national identity. Other studies have shown that language editions illustrate the same concepts with different images \cite{he2018the_tower_of_babel} and describe cultural artifacts in geographically proximate ways \cite{laufer2015mining}. 

Research on conflict-related content has also revealed ingroup bias across language editions. \citet{oeberst2020collectively} compared Wikipedia articles about 35 intergroup conflicts and found that each language edition portrayed its own group as less immoral and less responsible than the opposing group. Qualitative case studies have examined specific conflicts, including the Srebrenica massacre \cite{rogers2013wikipedia}, the Portuguese Colonial War \cite{ferreira2022construction}, and the Second Sino-Japanese War \cite{gustafsson2020international}, revealing persistent framing differences despite Wikipedia's neutrality norms.

Beyond content and framing, governance structures vary across language editions and can shape editorial outcomes \cite{hwang2022rules,johnson2022considerations,zhang2011group,hale2014multilinguals}. \citet{kharazian2024governance} showed that concentrated administrative power within Croatian Wikipedia enabled sustained nationalist revisionism, while the linguistically similar Serbian edition---with more distributed governance---did not exhibit comparable distortions.


\subsection{Connotation Frames for Narrative Analysis}

Prior work on Wikipedia bias has often relied on coverage-based measures or human-coded analyses of relatively small samples. In contrast, computational approaches enable a large-scale analysis of how narratives implicitly portray social actors and events. One such approach is \textit{connotation frames} \cite{rashkin2016connotation}, which capture the implied sentiment, power, and agency embedded in predicate--argument structures. Unlike explicit sentiment analysis, connotation frames focus on relational implications. For example, a verb such as ``defeated'' implies power and agency for its subject, whereas ``survived'' suggests vulnerability or reduced control for the agent.

Subsequent work has applied connotation frames to a variety of social and cultural contexts. \citet{sap2017connotation} used the framework to study gender bias in film scripts, showing that female characters are often portrayed with lower agency and power than male characters. \citet{rashkin2017multilingual} constructed multilingual connotation frame lexicons for 11 European languages and applied them to targeted sentiment analysis on social media. Most closely related to our work, \citet{park2021multilingual} analyzed Wikipedia biographies across multiple language editions and found systematic differences in how social groups are portrayed.


%% file: 03_ResearchQuestions.tex
\subsection{Research Questions}


Despite growing evidence of cross-linguistic variation in Wikipedia's coverage of conflict, prior work has primarily relied on qualitative case studies or small numbers of conflicts and language pairs, and has not systematically examined how combatant entities are relationally portrayed across a large set of wars and languages. Connotation frames offer a way to address this gap by capturing how power, agency, and sentiment are attributed through the relational structure of text, but have not yet been applied to war narratives. To address these gaps, we examine Wikipedia's coverage of wars through a cross-linguistic lens, investigating how language communities narrate conflicts differently based on their relationship to them. Specifically, we pose the following research questions:

\begin{itemize}[leftmargin=*]
    \item \textbf{RQ1:} How do Wikipedia language editions differ in their portrayal of combatant entities in wars with respect to power, agency, and sentiment, depending on their relationship to the war?
    \begin{itemize}
        \item \textbf{RQ1a:} When a language community is directly involved in a war, does that language portray entities aligned with itself differently from opposing entities?
        \item \textbf{RQ1b:} When multiple language communities are allied within the same war, how do involved languages portray entities associated with allied versus opposing camps?
    \end{itemize}
    \item \textbf{RQ2:} What topics characterize sentences in which self-aligned entities are framed as more versus less powerful or agentic across languages?
    \item \textbf{RQ3:} What factors help explain the observed cross-linguistic differences in how languages portray self and enemy entities in wars they are directly involved in?
    \item \textbf{RQ4:} When language communities are not directly involved in a war, do they still exhibit systematic differences in how they narrate the same conflict, or do their narratives converge toward similar structures?
\end{itemize}

%% file: 04_Data.tex
\section{Data Collection}
We collected all data through Wikipedia's official MediaWiki API\footnote{\scriptsize{\url{http://www.mediawiki.org/wiki/API:Main_page}}}. To identify the initial set of wars, we first collected all wars that occurred after 1900 listed under the English Wikipedia category ``List of wars by date.''\footnote{\scriptsize{\url{https://en.wikipedia.org/wiki/Category:Lists_of_wars_by_date}}} This category provides a tabular overview of major wars, including each war’s name, article link (English version), and the names and links of all belligerents. We removed wars in which the article link or any belligerent link was invalid, resulting in 788 wars. We also removed wars in which opposing belligerents shared at least one official language, as these cases prevent clean alignment between language communities and conflict sides. To determine each belligerent's official language(s), we linked entities to Wikidata and resolved their country affiliations (see Appendix \ref{appendix:language_assignment} for details). After removing all wars that contained belligerents with unknown or conflicting languages, 158 wars remained. 

To identify additional battle and conflict articles beyond the main war pages, we started from the Wikipedia articles for each of the 158 wars across all available language editions. We extracted all internal links from these articles and retained linked pages whose infoboxes listed belligerents and a valid conflict timeframe. We then applied the same language-assignment procedure to the belligerents of those pages and removed those with unknown or conflicting languages, yielding 1,558 additional pages. 

\begin{figure}[!htb]
    \centering
    \includegraphics[width=\linewidth]{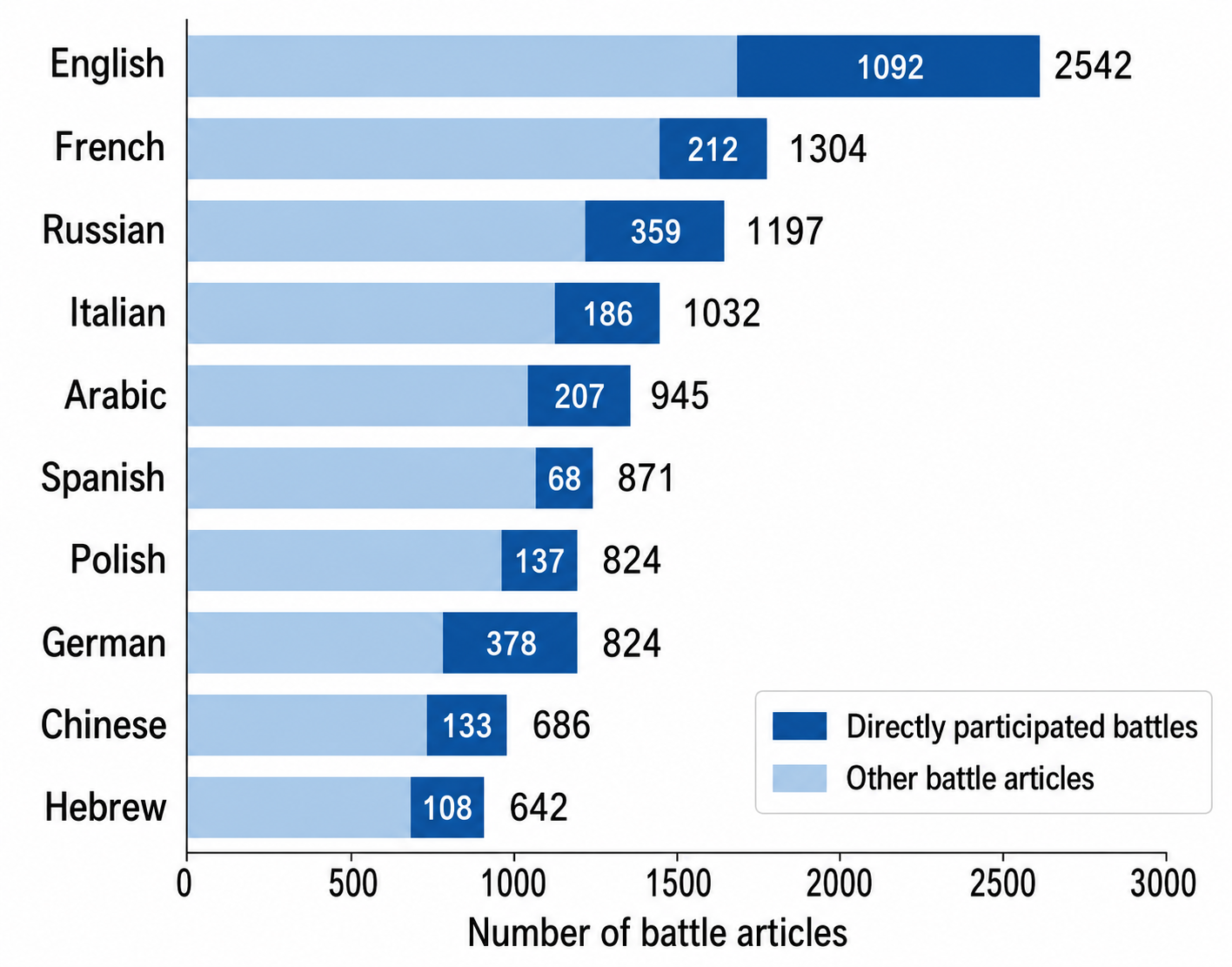}
    \caption{Distribution of battle articles across the top 10 languages in our dataset. Each bar represents the total number of battles available, with the darker segment indicating battles in which the language community was directly involved.}
    \label{fig:language-battle-distribution}
\end{figure}

For each war, we also collected its battle articles from the corresponding English Wikipedia category (e.g., ``\textit{Battles of Italo-Turkish War}''), whenever such a category existed. For World War I\footnote{\scriptsize{\url{https://en.wikipedia.org/wiki/List_of_military_engagements_of_World_War_I}}} and World War II \footnote{\scriptsize{\url{https://en.wikipedia.org/wiki/List_of_World_War_II_battles}}}, we instead used curated list articles to identify constituent battles. After de-duplicating and aggregating results from both the link-based expansion and the category-based collection, we obtained a final dataset of 2,581 battles. We then collected all available language editions and selected the 20 languages with the most articles and the highest number of battles in which the language community was directly involved---defined as battles whose infobox lists at least one belligerent using that language. Figure \ref{fig:language-battle-distribution} reports the distribution of the total number of available battle articles and the number of battles in which the language community directly participated.

We downloaded the full-text content of each article using the MediaWiki's TextExtracts API\footnote{\scriptsize{\url{https://www.mediawiki.org/wiki/Extension:TextExtracts}}}. We only kept the plain descriptive text of each article, excluding other types of information, such as tables, references, and infoboxes.

To ensure sufficient editorial activity in our analysis, we retained only articles with at least 10 edits \cite{holtz2018effects}. We chose this threshold because the 10th percentile of the edit count distribution is 11, so a threshold of 10 excludes only the least-edited articles. We verified that raising the threshold to higher values, such as 20 edits, would disproportionately reduce coverage of lower-resourced language editions such as Persian and Romanian. Applying the 10-edits criterion excluded 1,288 observations (7.4\% of the total sample), yielding a final sample of 16,058 articles across 20 languages and reducing the battle count from 2,581 to 2,542.

%% file: 05_Methodology.tex
\section{Methods}
\begin{figure*}[!htb]
    \centering
    \includegraphics[width=15cm, height=7.5cm]{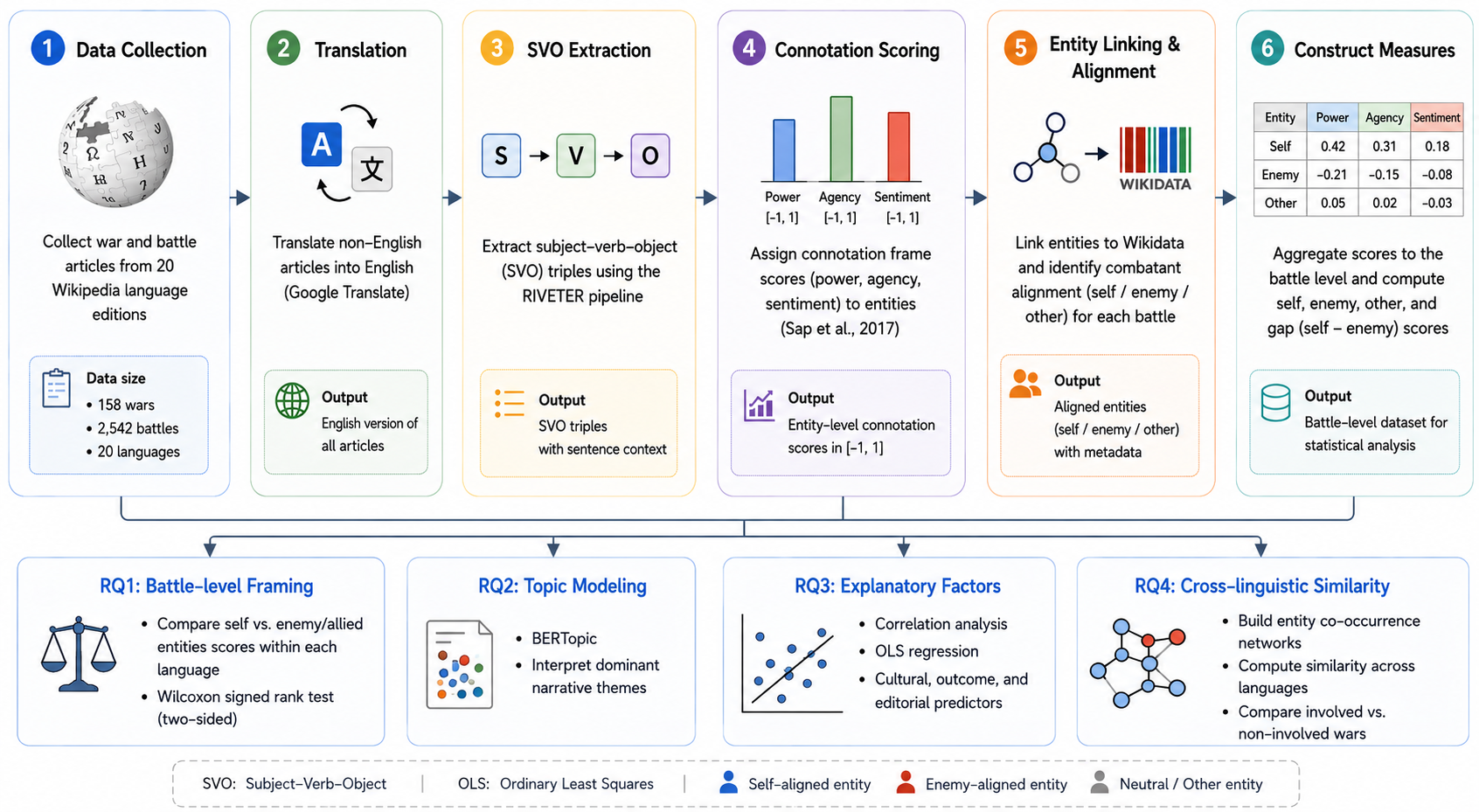}
    \caption{Overview of the six-stage method pipeline for analyzing cross-linguistic framing of wars and battles in Wikipedia.}
    \label{fig:pipeline}
\end{figure*}

Our methodology comprises six main stages (Figure \ref{fig:pipeline}): (1) collecting the data, (2) translating all articles into a common language, (3) extracting the subject-verb-object (SVO) from each sentence, (4) calculating entity-level connotation frames for power, agency, and sentiment, (5) linking entities to their national alignments, and (6) comparing narrative portrayals across languages. The scripts and materials needed to reproduce our analysis are available on OSF \cite{xia_gomez_zara_2026}.

\subsection{Translation}
After data collection, all non-English versions of articles were translated into English using the Google Translate API. To ensure that originally English articles did not receive systematically different connotation scores, we also back-translated them through Spanish, following established procedures for cross-linguistic comparability \cite{murayama2025linguistic,smith2022war}. To validate this translation pipeline, we replicated it using the Wikimedia Foundation's MinT translation service and obtained comparable results (see Appendix~\ref{appendix:translation} for more details).

\subsection{Connotation Frame}
We followed the RIVETER pipeline \cite{antoniak2023riveter}, which provides a comprehensive framework for extracting verb-centric connotation frames from text corpora. RIVETER integrates state-of-the-art NLP components---including dependency parsing, named-entity recognition, and coreference resolution---to identify agent–verb–theme triples and assign entities connotation scores based on its built-in lexicon of power, sentiment, and agency frames. These lexica encode directional relations for more than 1,700 frequently used verbs drawn from movie scripts and The New York Times articles.

In its original implementation, RIVETER outputs only entity-level aggregated average scores. We adapted this pipeline to retain each subject–verb–object (SVO) triple along with its associated connotation scores and a reference to its original sentence, enabling event-level comparisons within localized narrative contexts. Each verb was mapped to a categorical score of $+1$, $0$, or $-1$ for both the agent and theme across three dimensions: (1) \textit{Power}, which captures the implied social dominance or control in the event; (2) \textit{Agency}, which measures whether the agent is portrayed as acting with intentionality, initiative, or capacity to influence their environment, and (3) \textit{Sentiment}, which reflects the positive or negative affective presuppositions associated with the agent or theme. For example, in \textit{``Jack (agent) trapped Alice (theme)''}, Jack is scored as having more power and agency than Alice. 

To validate that there were no SVO classification differences across languages, we ran a human annotation study on Prolific in which native speakers of English, Arabic, Japanese, and Chinese evaluated 50 triples of SVO extractions from RIVETER in their respective languages. Agreement on power and agency dimensions did not differ significantly across languages, though sentiment showed lower consistency (see Appendix~\ref{appendix:translation} for more details).

\subsection{Entity Linking and Wikidata Enrichment}
To identify the national alignment of entities mentioned in battle narratives, we linked each entity to structured metadata from Wikidata and queried type-specific properties to infer country affiliation and geopolitical affiliation. For person entities, such as political leaders or military commanders, we used the ``country of citizenship'' property (\texttt{P27}), which captures formal national affiliation. When citizenship information was unavailable or insufficient, we additionally consulted the ``affiliation'' (\texttt{P1416}) and ``member of'' (\texttt{P463}) properties to capture relevant organizational or political alignments. For location and place-based entities---including cities, regions, and territories---we used the ``country property'' (\texttt{P17}). We further resolved national alignment through hierarchical containment relationships using ``located in the administrative territorial entity'' (\texttt{P131}) and ``part of'' (\texttt{P361}), which are particularly useful for historical regions or contested territories. For country entities, organizations, and military units, we followed the identification procedures described in the Data section. Entities for which no property yielded a confident country assignment were excluded.

\subsection{Analysis}
\subsubsection{Battle-Level Entity Portrayal Test.}
To answer RQ1a, we examined whether entities aligned with a language’s associated combatant were portrayed differently from opposing entities along the connotation frames. For each battle in each language edition, we computed average power, agency, and sentiment scores for each referenced entity using extracted SVO relations. Using entity linking results, we then categorized entities as \textit{self} (i.e., aligned with the combatant associated with the language edition) or \textit{enemy} (i.e., aligned with opposing combatants), and averaged scores within each category for each connotation dimension separately, yielding battle-level self and enemy scores for power, agency, and sentiment. This aggregation ensured that comparisons reflected the narrative portrayal of the same historical event rather than differences across conflicts. We tested for systematic differences between self and enemy scores using two-sided Wilcoxon signed-rank tests, implemented in \texttt{scipy} \cite{virtanen2020scipy}, as the distributions were non-normal and observations were naturally paired within each battle-language combination. To account for multiple comparisons, we applied the Benjamini–Hochberg false discovery rate correction \cite{thissen2002quick}. 

For RQ1b, we extended this analysis by categorizing entities as \textit{allied} (i.e., aligned with combatants on the same side as the focal language's country) and applied the same aggregation and testing procedure.

\subsubsection{Topic Modeling.}
To address RQ2, we divided sentences containing self-aligned SVO structures into two groups: those in which self-aligned entities were assigned higher power or agency scores, and those in which they were assigned lower scores. Using previously established alignment annotations, we mapped these SVOs back to their original sentences and performed topic modeling separately for each group within each language. We applied BERTopic \cite{grootendorst2022bertopic}, a sentence-level neural topic model well suited to capturing fine-grained semantic distinctions in localized narrative contexts. We chose BERTopic because our analysis compares localized sentence-level descriptions rather than full articles, making embedding-based clustering better suited than document-level topic models.

Prior to analysis, we removed topics dominated by named entities, such as personal names or country and nationality terms (e.g., Iran, Iraq, Iranian in Persian), as these topics primarily reflect referential specificity rather than narrative framing. For each language, we identified the most frequent topics associated with cases in which self-aligned entities received higher versus lower power or agency scores.

\subsubsection{Correlation Analysis.}
To address RQ3, we examined whether cultural values, editorial behaviors, and battle outcomes are associated with cross-linguistic differences in entity framing. We operationalized cultural values using Hofstede's five cultural dimensions: Power Distance, Individualism, Masculinity, Uncertainty Avoidance, and Long-Term Orientation \cite{hofstede2011dimensionalizing}. The Indulgence dimension was excluded due to missing values for several countries in our sample. Cultural scores were aggregated to the language level by averaging the country-level scores of all countries in which the language is listed as an official language.

We measured editorial behaviors using five indicators derived from Wikipedia battle articles: (a) \textit{revert rate}, computed as the proportion of edits that revert previous revisions within a language edition; (b) \textit{edit concentration}, calculated as the proportion of total edits contributed by the most active editors, reflecting how editing activity is distributed within a language community; (c) \textit{number of unique editors}, measured as the total count of distinct contributors; (d) \textit{talk page length}, measured as the total volume of talk page discussion content; and (e) \textit{section count}, captured by the number of sections in each article. The number of unique editors, talk page length, and section count were log-transformed prior to analysis to reduce right skew and stabilize variance. For the correlation analysis, these article-level indicators were aggregated to the language level by averaging across all battle articles within each language edition.

Battle outcomes were obtained from the infoboxes of English Wikipedia battle articles, with each battle coded as a win, loss, or draw (when no winner/loser was identified) for the focal combatant. For each language edition, we computed the win and loss rates as the proportions of battles in which the associated combatant was on the winning and losing sides, respectively.

Lastly, we operationalized narrative features as the language-level average of the connotation frame gap scores (i.e., power, agency, sentiment) derived from the RQ1 analysis, where each gap score reflects the difference between self and enemy entity scores within the same battle article. 

Given the small number of language editions ($n=20$), we conducted Spearman rank-order correlations between each combination of cultural values, editorial behaviors, win rate, and loss rate on the one hand, and average gap scores on the other. All correlations are reported as exploratory and should be interpreted with caution due to limited statistical power. To account for multiple comparisons, all reported significance levels were adjusted using the Benjamini--Hochberg correction procedure.

\subsubsection{Linear Regression Models.}
To examine whether framing differences persist after accounting for battle outcomes and editorial behaviors, we estimated ordinary least squares (OLS) regression models. We modeled three framing outcomes---self scores, enemy scores, and gap scores---across power, agency, and sentiment, yielding nine models. Each model included indicator variables for battle outcome (win or loss, with draw as the reference), the same editorial behavior controls used in the correlation analysis, computed at the article level, and language as a categorical covariate with English as the baseline (see Appendix~\ref{appendix:ols} for the full model specification). We standardized the continuous variables and checked normality before running the analysis. To reduce skewness, count-based variables were log-transformed. As a robustness check, we re-estimated each model using mixed-effects regression with language as a random intercept, and the results were consistent across specifications (see Appendix \ref{appendix:mixed-effects} for details).

\subsubsection{Network Similarity.}
To address RQ4, we compared narrative structures across languages using a graph-based similarity approach. For each battle in each language edition, we constructed an undirected, unweighted entity network, in which two entities are connected if they co-occur in the same SVO triple. Each node was assigned a categorical alignment label C1 (combatant 1), C2 (combatant 2), or OTHER (entities not belonging to either), and three continuous attributes representing its average power, agency, and sentiment scores within the battle.  We computed pairwise similarities between these networks using Propagation Kernels \cite{neumann2016propagation}, which capture structural similarities between graphs while natively supporting both discrete node labels and continuous node attributes. We computed these similarities using the GraKel Python package \cite{siglidis2020grakel}. 

We applied cosine normalization to control for differences in graph magnitude, then constructed an aggregated language–language similarity matrix by averaging all normalized similarity scores across battles in which a given language pair was jointly observed, excluding cases where either language was a direct participant. The resulting matrix was analyzed using agglomerative hierarchical clustering \cite{murtagh2012algorithms} to identify groups of languages with similar narrative patterns when describing wars in which they were not involved.

%% file: 06_Results.tex
\section{Results}
\subsection{RQ1: Framing Asymmetries in Wars Involving the Language Community}
\subsubsection{Self vs. Enemy Portrayals (RQ1a)}
Tables \ref{tab:self_greater_main} and \ref{tab:self_less_main} report the results of the Wilcoxon signed-rank tests comparing self-aligned and enemy-aligned entity scores across different connotation dimensions, corrected for multiple comparisons. Only languages exhibiting at least one significant result were included in the tables.

The results reveal divergent patterns across languages. Some language editions portray self-aligned entities as significantly more powerful or agentic. For example, English shows significantly higher agency for self-aligned entities, Japanese and Hebrew exhibit higher power and agency, and Russian exhibits significant differences across all three dimensions. In contrast, Arabic, Ukrainian, and Vietnamese portray self-aligned entities as significantly less powerful or agentic than enemy-aligned entities.

\input{tables/self_aligned_high}
\input{tables/self_aligned_low}

\subsubsection{Allied vs. Enemy Portrayals (RQ1b)}
The sample for this analysis was smaller than RQ1a, as not all battles included allied belligerent entities. Ukrainian and Hebrew are excluded from the statistical analysis due to insufficient sample sizes. After correcting for multiple comparisons, no language exhibited significant differences between allied and enemy portrayals on any dimension. This contrasts with the previous results, suggesting that narrative asymmetries in Wikipedia war articles primarily emerge in self-referential contexts rather than extending to broader alliance-based framing.

\subsubsection{Cross-linguistic Comparison of Verb Usage}
Since Russian was the only language with significant differences across all three dimensions, we examined its verb usage more closely. Comparing verbs associated with positive self-aligned descriptions revealed that \textit{award} appeared much more frequently in Russian (3\%) than in Japanese (0.02\%) or Hebrew (0.01\%). A permutation test confirmed that only Russian exhibited a significant asymmetry in \textit{award} usage, with the verb appearing more frequently in positive descriptions of self-aligned than enemy-aligned entities ($\Delta = 0.021$, $p = 0.003$).

This pattern reflects a distinctive feature of Russian war narratives, which frequently include retrospective commendation passages that record institutional recognition for bravery or sacrifice \cite{davis2017myth}. For example, the Russian account of the \textit{Battle of Robat Karim} states:
\begin{quote}
\textit{``At the decisive moment of the battle, he led his platoon in a bayonet charge and was killed in action. For his heroic deed, he was posthumously \textbf{awarded} the Cross of St. George.''}
\end{quote}
 


\begin{figure}[t]
    \centering
    \includegraphics[width=\linewidth]{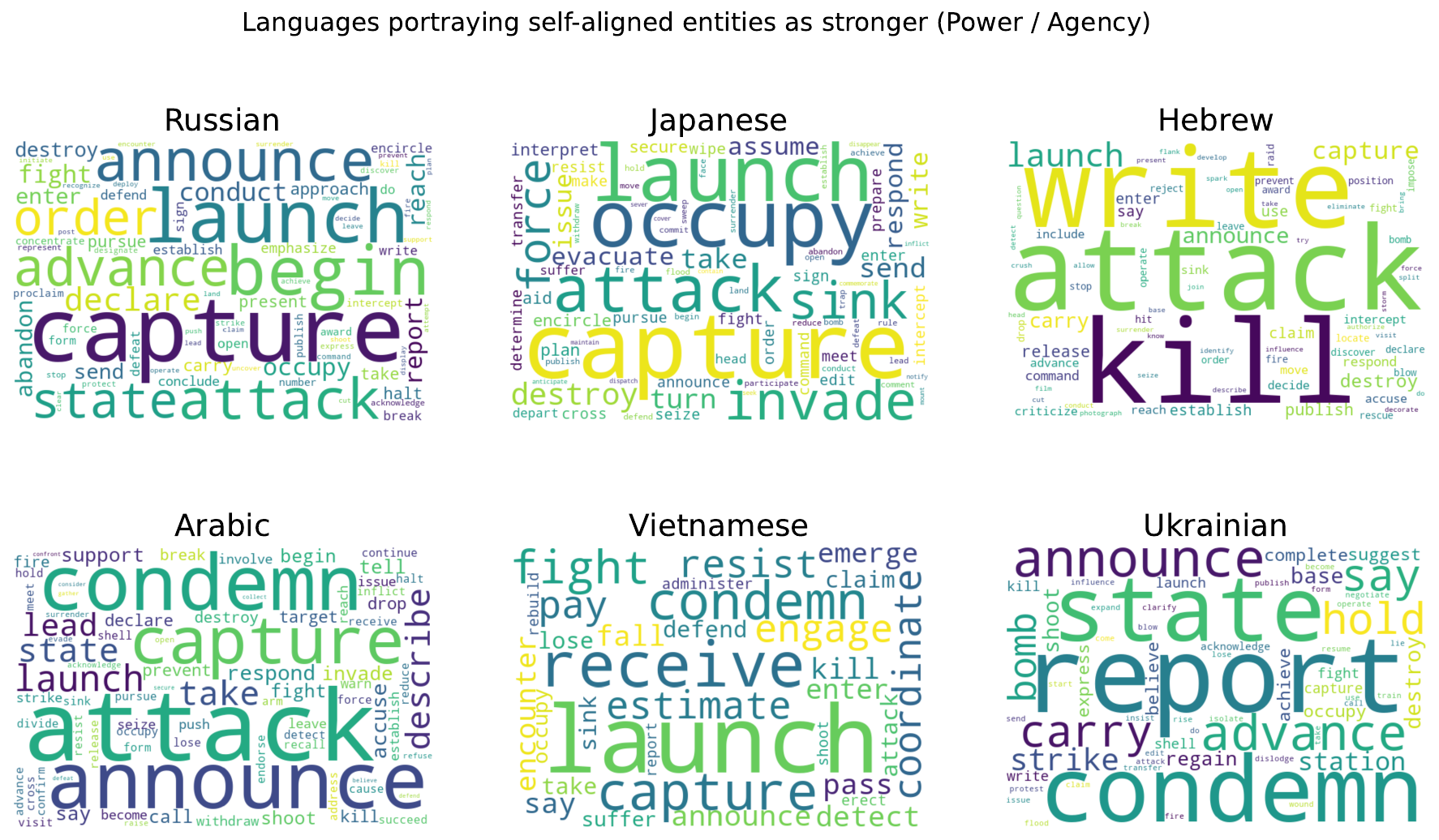}
    \caption{Verb word clouds for descriptions in which self-aligned entities receive \textit{higher} power and agency scores. Russian, Japanese, and Hebrew are shown in the top row, while Arabic, Vietnamese, and Ukrainian are shown in the bottom row.}
    \label{fig:verb-wordcloud-high}
\end{figure}

\begin{figure}[t]
    \centering
    \includegraphics[width=\linewidth]{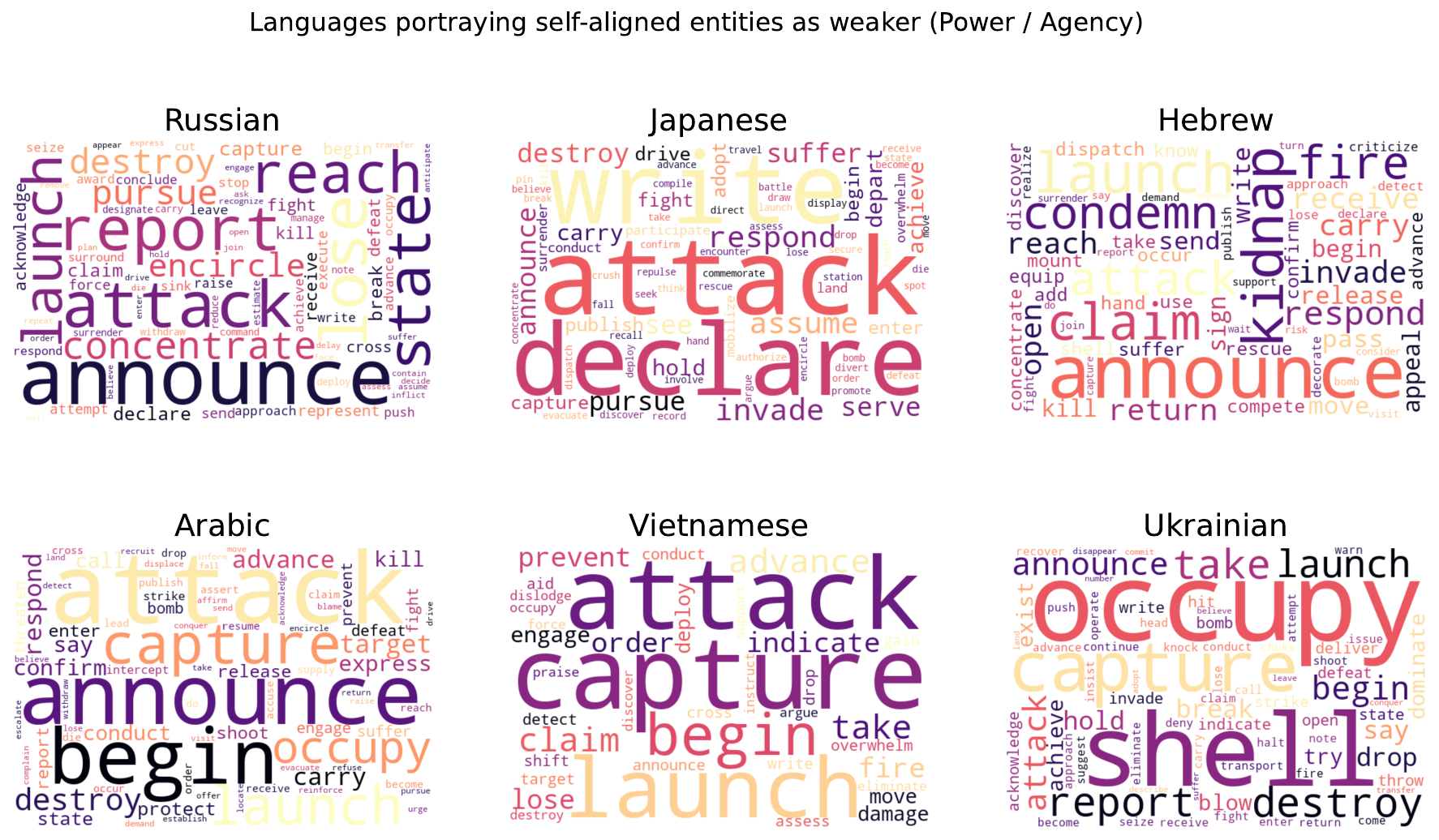}
    \caption{Verb word clouds for descriptions in which self-aligned entities receive \textit{lower} power and agency scores. While core war-action verbs appear across all languages, evaluative verbs such as \textit{condemn} are more prominent in these narratives.}
    \label{fig:verb-wordcloud-low}
\end{figure}

Beyond the Russian case, we compared verb usage across languages that portray self-aligned entities as stronger (Japanese, Hebrew, Russian) versus weaker (Arabic, Ukrainian, Vietnamese). Figures \ref{fig:verb-wordcloud-high} and \ref{fig:verb-wordcloud-low} present the resulting word clouds for sentences in which self-aligned entities are assigned higher and lower power or agency, respectively. Core military verbs such as \textit{attack}, \textit{occupy}, and \textit{launch} appear frequently regardless of the self-identity framing. However, the verb \textit{condemn} shows a distinct cross-linguistic pattern. It appears more frequently in Arabic, Vietnamese, and Ukrainian when self-aligned entities are framed as stronger or more agentic, whereas in Hebrew, it primarily appears in contexts where self-aligned entities receive lower scores. In these cases, \textit{condemn} frames the self-aligned actor through moral responses rather than direct military action, as illustrated in the Arabic version of the \textit{Al-Fashaga conflict}:

\begin{quote}
\textit{``On January 13, an Ethiopian Air Force aircraft violated Sudanese airspace, the Sudanese Foreign Ministry \textbf{condemned} the incident.''}
\end{quote}

Here, the verb \textit{condemned} frames the self-aligned actor through moral response to an adversary’s behavior. 

\subsection{RQ2: Thematic Differences in Self-Aligned Framing}
The topic modeling results (Table \ref{tab:rq2-topics-wide}) reveal a systematic contrast between languages that portray self-aligned entities as relatively weaker and those that portray them as stronger. Languages such as Arabic, Ukrainian, and Vietnamese---where self-aligned entities tend to receive lower power or agency scores---exhibit a pronounced resistance-and-defense framing when self-entity scores are higher, characterized by verbs such as \textit{resist}, \textit{defend}, and \textit{protect}. When self-entity scores are lower, these languages, instead, emphasize verbs denoting failure or inability (e.g., fail, unable). This pattern may suggest a narrative structure characterized by \textit{``we resisted, but ultimately failed.''}

In contrast, languages that portray self-aligned entities as stronger do not display a consistent narrative focus when self-entity scores are higher. However, when self-entity scores are lower, these languages consistently emphasize in-group casualties and losses, foregrounding the costs borne by the in-group. Languages that exhibit more neutral self-entity portrayals, such as Chinese and Spanish, do not show a clear or consistent narrative pattern across conditions.

\input{tables/topic_modeling}

\subsection{RQ3: Predictors of Cross-Linguistic Framing Differences}

\subsubsection{Correlation Analysis}
\begin{figure}[!htb]
    \centering
    \includegraphics[width=\linewidth]{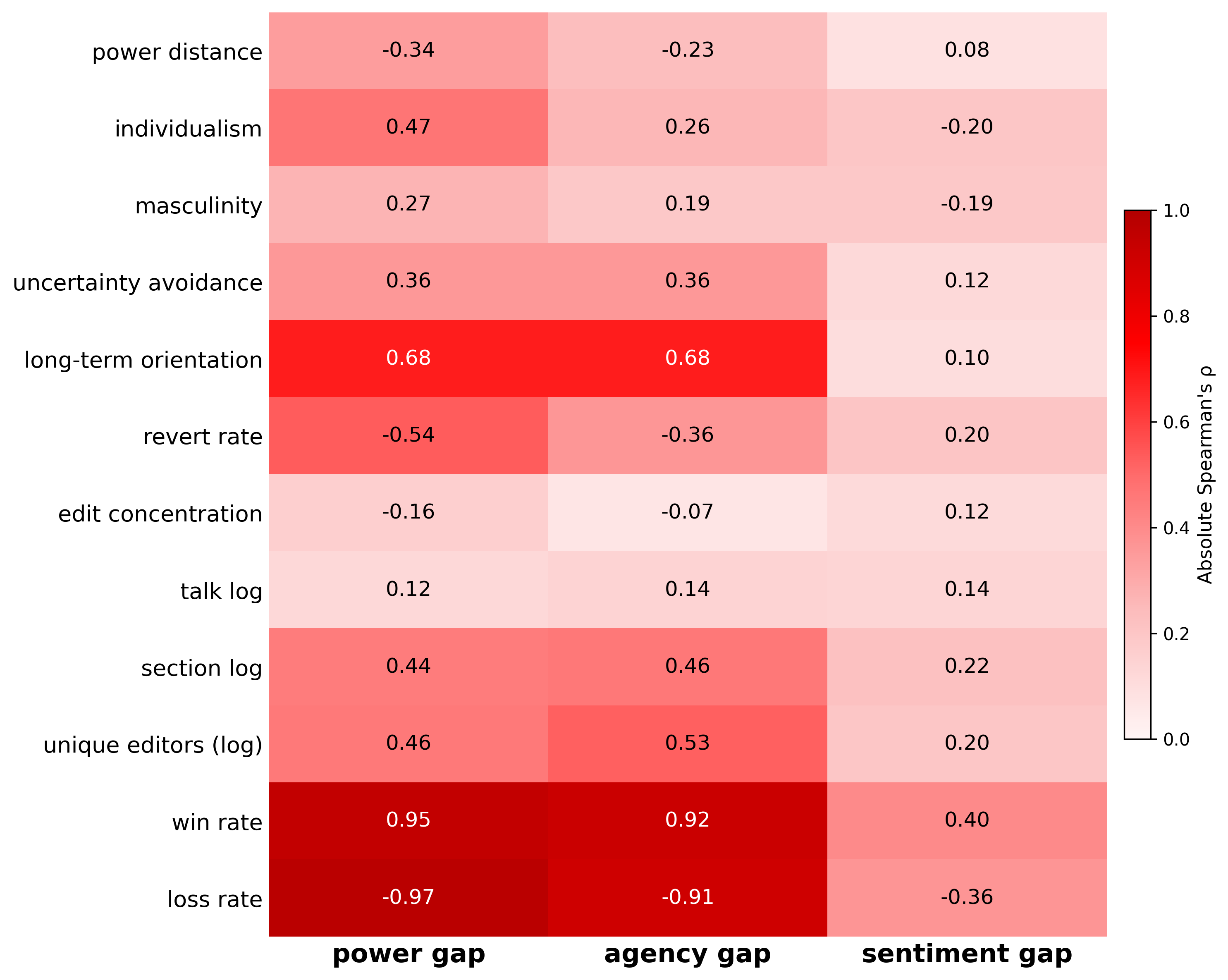}
    \caption{Spearman correlation heatmap between cultural, editorial, and battle-outcome variables and self--enemy framing gaps in power, agency, and sentiment. Cell values report Spearman's $\rho$, with darker shading indicating stronger absolute correlations.}
    \label{fig:correlation_heatmap}
\end{figure}

Three variables showed significant associations with both the power gap and the agency gap (Figure \ref{fig:correlation_heatmap}). Win rate was strongly positively correlated with the power gap ($\rho = .95$, $p_{adj} < .001$) and agency gap ($\rho = .92$, $p_{adj} < .001$), whereas loss rate showed strong negative correlations with power gap ($\rho = -.94$, $p_{adj} < .001$) and agency gap ($\rho = -.91$, $p_{adj} < .001$). Long-Term Orientation showed moderate positive correlations with power gap ($\rho = .68$, $p_{adj} < .01$) and agency gap ($\rho = .68$, $p_{adj} < .01$), suggesting that cultures prioritizing perseverance and long-term planning may tend toward stronger self-aligned framing in war narratives. No significant correlations were observed for the sentiment gap, nor for the remaining cultural dimensions or editorial behavior variables.

Overall, languages with higher proportions of victories and stronger long-term orientation tendencies tended to exhibit larger self-enemy gaps in power and agency portrayals, whereas languages with higher loss rates tended to exhibit smaller gaps. For instance, the Japanese Wikipedia has one of the highest win rates (61\%) and the highest Long-Term Orientation score (100), alongside the largest power and agency gaps, while Arabic and Vietnamese Wikipedia show lower win rates (31\% and 32\%), lower Long-Term Orientation scores (25 and 34), and substantially smaller gaps.

\subsubsection{Linear Regression Models} 
The results reveal consistent framing patterns across language communities. Japanese, Hebrew, and Russian articles exhibit significantly higher self-framing scores in both power and agency than English articles, indicating a clear tendency toward self-enhancement. Japanese articles not only portray self-aligned entities as more powerful and agentic, but also tend to reduce portrayals of enemy power. Russian articles further describe self-aligned entities more positively in the sentiment dimension, which is consistent with the patterns observed in RQ1. Because these languages systematically elevate portrayals of self-aligned actors---in some cases, simultaneously weakening portrayals of opposing actors---they ultimately produce significantly larger framing gaps than English does.

Arabic and Vietnamese display the opposite pattern, portraying self-aligned entities as weaker and less agentic while simultaneously depicting enemy entities as stronger or more agentic, resulting in significantly smaller framing gaps. Ukrainian and Persian articles primarily differ from English through stronger portrayals of enemy entities rather than through weaker self-portrayals. German exhibits a more nuanced pattern: its articles significantly increase self-oriented power portrayals, but also convey more positive sentiment toward enemy entities, suggesting that framing dimensions do not necessarily move in the same direction within a language community.

Battle outcomes were consistently associated with framing differences across multiple dimensions. Victories were associated with significantly larger power, agency, and sentiment gaps, driven primarily by reduced portrayals of enemy entities. Losses were linked to smaller power and agency gaps, partially through reduced self-oriented power and agency portrayals, while in some cases also increasing enemy power portrayals.

Several editorial behavior variables were also associated with framing outcomes. Higher edit concentration was associated with larger agency gaps and lower enemy agency. This finding suggests that articles dominated by a smaller group of editors may be more likely to suppress agentic portrayals of opposing actors. Higher revert rates were negatively associated with enemy-agency portrayals, indicating that greater editorial conflict was associated with weaker depictions of enemy actors. In addition, articles with more unique editors tended to exhibit higher self- and enemy-power scores and more positive portrayals of enemy sentiment. This result suggests that broader participation may contribute to more elaborate or intensified descriptions of conflict actors. Longer article structures also showed selective associations with framing. Greater section counts were negatively associated with self-agency portrayals, indicating that more structurally detailed articles may adopt more complex narrative styles.

\input{tables/ols_regression}

\subsection{RQ4: Narrative Convergence in Non-Involved Languages}
\begin{figure}[t]
    \centering
    \includegraphics[width=\linewidth]{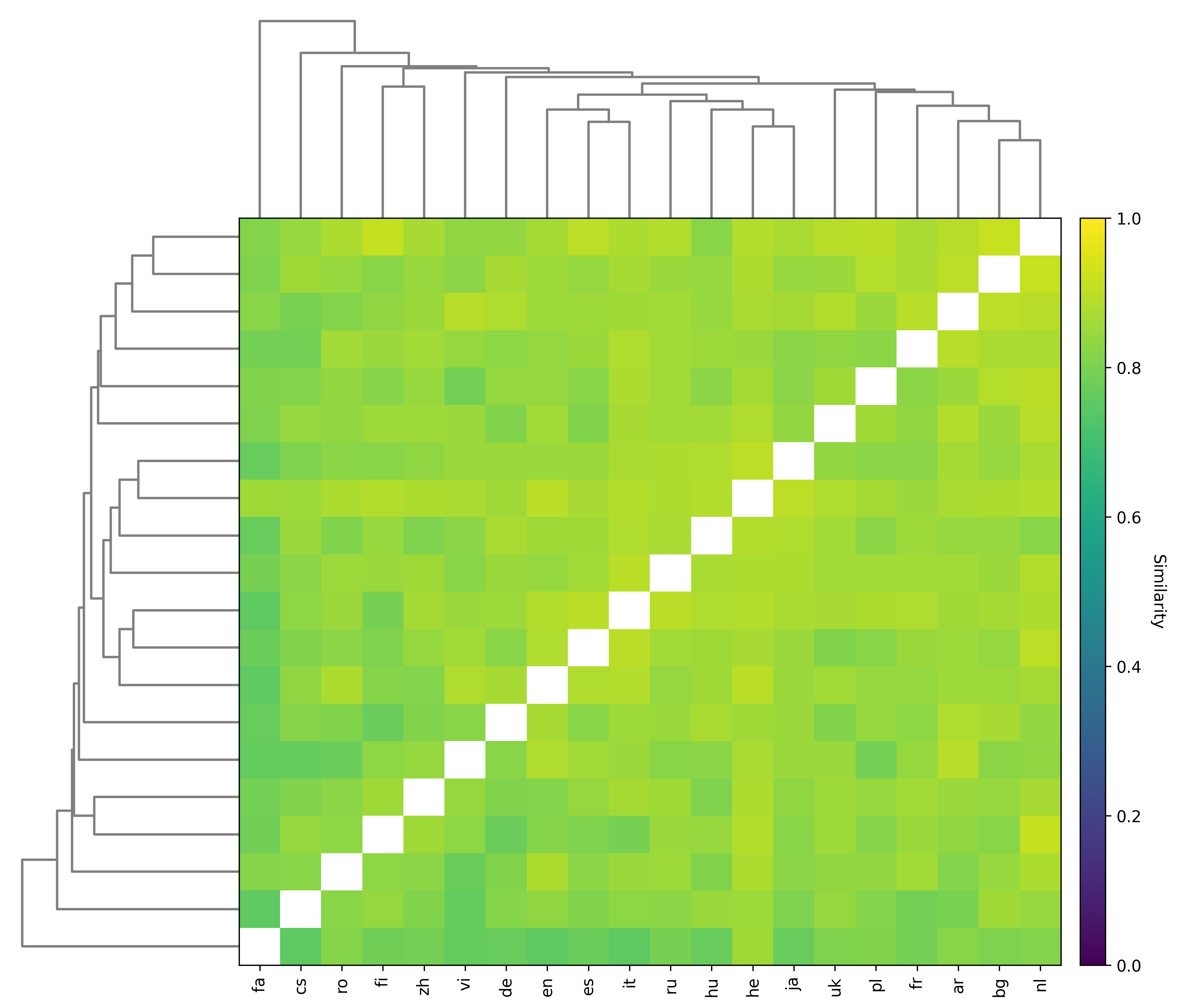}
    \caption{Clustered heatmap with hierarchical dendrograms showing narrative similarity between languages in battles in which they are not directly involved. Similarity values range from 0.7 to 0.9, and the diagonal is suppressed to emphasize cross-language relationships.}
    \label{fig:non_involved_similarity}
\end{figure}

To address RQ4, Figure \ref{fig:non_involved_similarity} presents a clustered heatmap with hierarchical dendrograms showing narrative similarity between languages when describing wars in which they are not directly involved. Similarity scores are uniformly high, ranging from 0.7 to 0.9, and the dendrogram shows no clear separations indicative of discrete clusters. Despite this high overall similarity, the dendrogram reveals differences in positioning. Several languages, such as Arabic and Dutch, consistently appear closer to the center of the similarity space. In contrast, languages including Persian, Czech, Vietnamese, and Finnish tend to occupy more peripheral positions, branching off earlier in the dendrogram. 

%% file: tables/self_aligned_high.tex
\begin{table}[!htb]
\centering
\small
\begin{tabular}{llrr}
\hline
\textbf{Language} & \textbf{Dimension} & \textbf{$p_{adj}$} & \textbf{Sig.} \\
\hline
\multirow{1}{*}{English}
 & Agency  & $<0.001$ & *** \\
\hline
\multirow{2}{*}{Hebrew}
 & Power  &  0.007 & ** \\
 & Agency &  $<0.001$ & *** \\
 \hline
\multirow{2}{*}{Japanese}
 & Power &  $<0.001$ & *** \\
 & Agency  & 0.002 & ** \\
\hline
\multirow{3}{*}{Russian}
 & Power  &  0.004 & ** \\
 & Agency  &  0.002 & ** \\
 & Sentiment  & 0.018 & * \\
\hline
\end{tabular}
\caption{Wilcoxon signed-rank test results for languages in which self-aligned entities are portrayed as \textit{more} powerful, \textit{more} agentic, or \textit{more positive} than enemy-aligned entities. Benjamini--Hochberg false discovery rate correction was applied. $p$-values are two-sided. Significance levels: *** $p<0.001$, ** $p<0.01$, * $p<0.05$.}
\label{tab:self_greater_main}
\end{table}

%% file: tables/self_aligned_low.tex
\begin{table}[!htb]
\centering
\small
\begin{tabular}{llrr}
\hline
\textbf{Language} & \textbf{Dimension} & \textbf{$p_{adj}$} & \textbf{Sig.} \\
\hline
\multirow{2}{*}{Arabic}
 & Power  &  $<0.001$ & *** \\
 & Agency  &  $<0.001$ & *** \\
\hline
\multirow{2}{*}{Ukrainian}
 & Power   & 0.004 & ** \\
 & Agency  & 0.010 & ** \\
\hline
\multirow{2}{*}{Vietnamese}
 & Power  &  0.006 & ** \\
 & Agency  & 0.005 & ** \\
\hline
\end{tabular}
\caption{Wilcoxon signed-rank test results for languages in which self-aligned entities are portrayed as \textit{less} powerful, \textit{less} agentic, or \textit{more negative} than enemy-aligned entities.
Significance levels: *** $p<0.001$, ** $p<0.01$, * $p<0.05$. }
\label{tab:self_less_main}
\end{table}

%% file: tables/topic_modeling.tex
\begin{table*}[!htb]
\centering
\small
\setlength{\tabcolsep}{4pt}
\renewcommand{\arraystretch}{1.2}
\begin{tabular}{lcccc}
\hline
 & \textbf{Arabic} & \textbf{Persian} & \textbf{Ukrainian} & \textbf{Vietnamese} \\
\hline
High score
& defense, defend, protect
& forces, fight, counterattack
& resist, counter, defend
& defend, defensive, guard\\
Low score
& impossible, hamper, unable
& suffered, destroyed, casualties
& due, lack, failure
& fail, difficult, struck\\
\hline
 & \textbf{Japanese} & \textbf{Hebrew} & \textbf{Russian} & \textbf{Italian} \\
\hline
High score
& capture, advanced, attacked
& shoot, kill, fire
& armed, corps, guns
& command, fight, weapons\\
Low score
& killed, fierce, casualties
& losses, lost, missing
& wound, losses, suffer
& total, suffered, killed\\
\hline
 & \textbf{Chinese} & \textbf{Polish} & \textbf{Romanian} & \textbf{Spanish} \\
\hline
High score
& hit, damaged, causing
& defense, forces, uprising
& weapons, armed, war
& united, fortified, attacked\\
Low score
& advanced, offensive, attacked
& total, lost, dead
& bombers, fighters, defensive
& hampered, bombs, advance\\
\hline
 & \textbf{Dutch} & \textbf{Czech} & \textbf{Hungarian} & \textbf{English} \\
\hline
High score
& constructed, defensive, front
& division, army, general
& shot, bomber, fight
& force, shot, fight\\
Low score
& failed, unable, not
& wounded, lost, missing
& casualties, losses, suffered
& wounded, dead, total\\
\hline
 & \textbf{Bulgarian} & \textbf{Finnish} & \textbf{German} & \textbf{French} \\
\hline
High score
& achieved, war, military
& guards, defensive, protect
& corps, army, guards
& defensive, heavy, repelled\\
Low score
& fire, hit, damaged
& attacked, capture, advanced
& struck, destroyed, damaged
& wounded, killed, dead\\
\hline
\end{tabular}
\caption{Top topic keywords associated with self-aligned entities under higher versus lower power/agency scores across 20 languages. For each language, we report the three most representative keywords of the most frequent topic under each condition.}
\label{tab:rq2-topics-wide}
\end{table*}

%% file: tables/ols_regression.tex
\begin{table*}[!htb]
\centering
\scriptsize
\setlength{\tabcolsep}{4pt}
\renewcommand{\arraystretch}{1}

\begin{tabular}{lccccccccc}
\hline

& \multicolumn{3}{c}{\textbf{Power}}
& \multicolumn{3}{c}{\textbf{Agency}}
& \multicolumn{3}{c}{\textbf{Sentiment}} \\

\cline{2-10}

\textbf{Predictor}
& \textbf{Self}
& \textbf{Enemy}
& \textbf{Gap}
& \textbf{Self}
& \textbf{Enemy}
& \textbf{Gap}
& \textbf{Self}
& \textbf{Enemy}
& \textbf{Gap} \\

\hline

\multicolumn{10}{l}{\textit{Outcome \& Editorial Behaviors}} \\
\hline

Win
& ---
& -0.091***
& +0.114**
& ---
& -0.077***
& +0.076**
& ---
& -0.077***
& +0.102*** \\

Loss
& -0.092***
& +0.074**
& -0.166***
& -0.094***
& ---
& -0.125***
& ---
& -0.034.
& --- \\

Revert rate (z)
& ---
& ---
& ---
& ---
& -0.016*
& ---
& ---
& ---
& --- \\

Edit concentration (z)
& ---
& ---
& ---
& ---
& -0.029***
& +0.023*
& ---
& -0.016*
& --- \\

Log unique editors (z)
& +0.032**
& +0.028*
& ---
& ---
& ---
& ---
& ---
& +0.020*
& --- \\

Section log (z)
& ---
& ---
& ---
& -0.023**
& ---
& -0.024.
& +0.023**
& +0.023**
& --- \\

Talk log (z)
& ---
& ---
& ---
& +0.015.
& ---
& +0.020.
& ---
& ---
& +0.018. \\

\hline

\multicolumn{10}{l}{\textit{Language Effects (ref: English)}} \\
\hline

Japanese
& +0.252***
& -0.085.
& +0.336***
& +0.164***
& ---
& +0.167***
& ---
& +0.103**
& -0.103* \\

Hebrew
& +0.250***
& ---
& +0.263**
& +0.155***
& ---
& +0.141*
& ---
& ---
& --- \\

Russian
& +0.188***
& ---
& +0.178***
& +0.110***
& ---
& +0.103**
& +0.083***
& ---
& +0.056. \\

German
& +0.135***
& ---
& +0.117*
& +0.039.
& ---
& ---
& ---
& +0.097***
& -0.134*** \\

Arabic
& -0.123**
& +0.139**
& -0.262***
& -0.085**
& +0.112***
& -0.197***
& ---
& +0.094**
& -0.130*** \\

Ukrainian
& ---
& +0.143*
& -0.205*
& ---
& +0.136**
& -0.181**
& ---
& ---
& --- \\

Persian
& ---
& +0.315**
& -0.424*
& ---
& +0.282***
& -0.285*
& ---
& ---
& --- \\

Vietnamese
& -0.174*
& +0.229**
& -0.403**
& ---
& +0.185**
& -0.259**
& ---
& +0.115.
& --- \\

Hungarian
& ---
& ---
& ---
& -0.195***
& ---
& -0.257***
& -0.085.
& +0.160**
& -0.245*** \\

Dutch
& ---
& +0.113
& -0.115
& -0.114*
& ---
& -0.135.
& ---
& +0.164**
& -0.213** \\

Italian
& +0.132**
& ---
& +0.123.
& ---
& ---
& ---
& ---
& +0.108***
& -0.088* \\

Spanish
& ---
& ---
& ---
& ---
& ---
& ---
& ---
& +0.117*
& -0.128* \\

French
& +0.081*
& ---
& +0.102.
& ---
& ---
& ---
& ---
& +0.067*
& --- \\

Chinese
& ---
& +0.095.
& ---
& ---
& +0.097**
& ---
& ---
& ---
& -0.091. \\

Bulgarian
& +0.193**
& +0.100
& ---
& +0.107*
& +0.099.
& ---
& ---
& ---
& --- \\

Romanian
& ---
& +0.135.
& -0.190.
& ---
& ---
& ---
& ---
& +0.104.
& --- \\

Czech
& ---
& -0.267*
& +0.392*
& ---
& ---
& ---
& ---
& ---
& --- \\

\hline

Residual SD
& .4946 & .5408 & .7808
& .3669 & .3822 & .5368
& .3652 & .3807 & .4874 \\

$R^2$
& .156 & .133 & .157
& .151 & .139 & .160
& .124 & .130 & .131 \\

Adj. $R^2$
& .149 & .126 & .151
& .145 & .132 & .153
& .117 & .123 & .124 \\

\hline
\multicolumn{10}{l}{\scriptsize Significance: *** $p<.001$, ** $p<.01$, * $p<.05$, . $p<.10$} \\

\hline
\end{tabular}
\caption{
OLS regression results for self-, enemy-, and gap-framing scores across power, agency, and sentiment dimensions 
(reference outcome: draw; reference language: English; gap score = self score $-$ enemy score). 
Continuous editorial predictors were standardized before estimation. Residual SD denotes the residual standard deviation of the OLS model.
}

\label{tab:combined_framing_models}

\end{table*}

%% file: 07_Discussion.tex
\section{Discussion}
This study provides a large-scale cross-linguistic analysis of how Wikipedia language editions narrate wars. Our findings show that narrative differences are pronounced when language communities describe battles in which they are directly involved, and these differences are associated with war outcomes and editorial dynamics. When languages describe battles in which they are not directly involved, narratives largely converge, suggesting that direct involvement is an important source of cross-linguistic divergence.

Our findings extend and complicate prior work on ingroup bias. While \citet{oeberst2020collectively} showed that language editions portray their own group as less immoral and responsible, we find that the direction of bias is not uniform: English, Japanese, Hebrew, and Russian portray self-aligned entities as more powerful and/or agentic, whereas Arabic, Ukrainian, and Vietnamese portray them as less so. Moreover, these asymmetries are self-referential rather than alliance-based. No language exhibited significant differences between allied and enemy portrayals, providing no evidence that these framing asymmetries extend to broader alliance-based identification. The direction of bias also corresponds to distinct narrative structures: languages that portray self-aligned entities as weaker tend to emphasize resistance followed by failure, while those that portray them as stronger foreground in-group casualties and losses. Multilingual framing differences thus cannot be reduced to a single form of ingroup favoritism.

Our exploratory correlation analysis of the 20 languages identified two variables consistently associated with power and agency gaps: win rate and long-term orientation. Languages with higher proportions of victories and higher long-term orientation scores tended to exhibit larger self-enemy gaps, though these associations should be interpreted with caution, given the small sample size. The regression models, which operate at the article level, corroborate the role of battle outcomes. In summary, victories were associated with larger framing gaps, driven primarily by reduced portrayals of enemy entities, while losses were linked to smaller gaps through reduced self-portrayals. Importantly, the regression models also revealed that similar framing gaps can emerge through different narrative mechanisms. For example, Japanese articles enlarge gaps by both elevating self-portrayals and suppressing enemy portrayals, whereas Hebrew and Russian articles produce comparable gaps primarily through stronger self-portrayals alone.

These findings also carry important practical implications for Wikipedia communities. RQ3 shows that higher editorial concentration is associated with suppressed agentic portrayals of opposing entities, suggesting that unequal distributions of editorial influence shape how conflicts are narratively constructed, quantitatively complementing prior qualitative research on governance capture in specific language editions \cite{kharazian2024governance}. These patterns suggest that promoting editorial heterogeneity over politically sensitive topics may contribute to more balanced narratives, though causal claims require more investigation.

RQ4 offers a particularly actionable implication. When editors revise articles about wars in which their own community was directly involved, they may benefit from consulting language editions not directly implicated in the conflict, which, in our analysis, exhibited high cross-linguistic similarity and may serve as an additional comparative reference. This finding could inform the design of editorial assistance tools. For example, Wikipedia could surface entity descriptions from non-involved language editions, helping editors recognize potential framing asymmetries while preserving the editorial autonomy of language communities.

Our findings nuance common critiques of Wikipedia's neutrality \cite{oeberst2016individual,kharazian2024governance} by showing that narrative divergence across language editions is conditional rather than uniform. The application of neutrality norms appears shaped by editors' cultural and linguistic proximity to the subject matter, particularly in controversial domains such as war. Wikipedia's multilingual coverage thus emerges from an interaction between shared editorial norms and locally situated perspectives. 

\subsection{Limitations and Future Work} 
We acknowledge the limitations of this study. First, our analysis relies on translating all articles into English, which may obscure subtle semantic distinctions and culturally specific connotations. Building language-specific connotation lexicons for all 20 languages was infeasible at this scale. To assess translation robustness, we conducted a human validation study evaluating SVO extraction, verb translation quality, and connotation-frame judgments across languages, which demonstrated acceptable agreement on the power and agency dimensions (see Appendix \ref{appendix:translation}). However, sentiment scores showed consistently lower inter-annotator agreement across languages and should be interpreted with particular caution. Moreover, the use of Spanish as an intermediate language in back-translation may introduce its own distortions, and we cannot fully rule out the possibility that some cross-linguistic differences reflect translation artifacts rather than genuine narrative variation.

Second, the number of battles varies substantially across language editions, which may affect statistical power for smaller language communities, such as Persian and Romanian. Third, our correlation analysis is limited by the small sample of 20 languages. The number of battles in which each language community directly participated also varies substantially across languages (e.g., 1,092 for English versus 108 for Hebrew), which may reduce statistical power for languages with fewer observations. Relatedly, articles with sparse edit histories may show higher narrative convergence simply because there is less room for divergence. Although we applied a minimum threshold of 10 edits, we did not stratify results by edit volume.

Fourth, our analysis is descriptive and comparative rather than causal: we document cross-linguistic differences in narrative framing but cannot make causal claims about the mechanisms producing them, as observed patterns may reflect editor composition, source availability, or cultural memory traditions that our data cannot fully disentangle. 

Fifth, we did not account for variation in the geographic distribution of language speakers. Some languages in our sample (e.g., English, Spanish) are widely adopted across multiple countries with diverse geopolitical positions, while others (e.g., Hebrew, Vietnamese) are more concentrated in specific national contexts. This heterogeneity may complicate the interpretation of language-level patterns, as editions associated with geographically diffuse languages aggregate editors with potentially divergent national perspectives. Additionally, our binary classification of entities as self-aligned or enemy-aligned may oversimplify conflicts involving multiple factions, non-state actors, or civil wars where allegiances are fluid or internally contested.

Lastly, we did not conduct a human evaluation of the pipeline's overall correctness for the whole corpus. Errors in dependency parsing, coreference resolution, or entity identification may affect connotation scores in ways that our validation did not capture. Our Prolific validation study confirmed that error propagation was comparable across languages for agency and power, but not for sentiment. 

Future work should examine conflicts in which opposing sides share a common language to determine whether similar framing patterns arise when linguistic communities overlap. It would also be valuable to analyze ostensibly factual elements of articles, such as casualty counts, timelines, or territorial boundaries, to assess whether cross-linguistic differences extend beyond narrative framing into the presentation of quantitative information. Future work could also incorporate larger-scale human evaluation of the NLP pipeline across additional languages to assess absolute accuracy beyond cross-linguistic comparability. Moreover, it could examine temporal dynamics, such as whether framing patterns shift around conflict milestones or changes in editorial protection policies. Finally, this computational pipeline could be applied to other contested topics on Wikipedia, helping identify the conditions under which multilingual knowledge production diverges across communities and when it converges toward shared narratives.

\section{Conclusion}
This study demonstrates that Wikipedia's war narratives diverge systematically across language editions when communities describe conflicts in which they are directly involved, but largely converge when they do not. These patterns extend prior findings on ingroup bias by showing that the direction and mechanisms of framing asymmetries vary across languages, and that we find these asymmetries for self-referential identification but not for broader alliance structures. By applying connotation frames at scale across 20 language editions, we move beyond coverage-based metrics to reveal how power, agency, and sentiment are differentially attributed to combatant entities. Our findings suggest that Wikipedia's neutrality may be best understood not as a fixed property of its policies, but as an emergent and conditional outcome of the relationship between editorial communities and the events they describe.

\section{Acknowledgement}
We thank the Google Cloud Research Credits Program for providing computing credits that supported the translation component of this work. We also thank the anonymous reviewers for their constructive feedback and Dr. Shiran Dudy for her helpful comments on the manuscript.

\section{Generative AI Statement}
Generative AI tools were used to support manuscript revision and to assist in understanding and interpreting the research results.

%% file: 08_EthicsChecklist.tex
\subsection{Paper Checklist to be included in your paper}

\begin{enumerate}
\item For most authors.../-+
\begin{enumerate}
    \item Would answering this research question advance science without violating social contracts, such as violating privacy norms, perpetuating unfair profiling, exacerbating the socio-economic divide, or implying disrespect to societies or cultures?
    \answerYes{Yes. This research analyzes publicly available Wikipedia articles and does not involve private user data.}
  \item Do your main claims in the abstract and introduction accurately reflect the paper's contributions and scope?
    \answerYes{Yes. The abstract and introduction state that we conduct a descriptive, comparative analysis of cross-linguistic differences in war narratives. We do not overclaim causal relationships.}
   \item Do you clarify how the proposed methodological approach is appropriate for the claims made? 
    \answerYes{Yes.  We explain why connotation frames are well-suited for capturing relational portrayals of entities, justify our use of the RIVETER pipeline, and describe our entity-linking and statistical testing procedures.}
   \item Do you clarify what are possible artifacts in the data used, given population-specific distributions?
    \answerYes{Yes. We acknowledge that the number of available battles varies across language editions, that some languages have broader geographic distributions of languages than others, and that translation may obscure language-specific nuances.}
  \item Did you describe the limitations of your work?
    \answerYes{Yes. The Limitations section addresses translation artifacts, sample size variation, small language sample, lack of human evaluation, the descriptive (non-causal) nature of our analysis, and variation in geographic distribution of language speakers.}
  \item Did you discuss any potential negative societal impacts of your work?
    \answerYes{Yes. We acknowledge that findings about systematic differences in how language communities portray war could potentially be misused to delegitimize Wikipedia as a knowledge source or to inflame intergroup tensions.}
  \item Did you discuss any potential misuse of your work?
    \answerYes{Yes. We note in the Limitations the potential for misuse and clarify that our aim is to increase transparency about how collective memory is constructed.}
  \item Did you describe steps taken to prevent or mitigate potential negative outcomes of the research, such as data and model documentation, data anonymization, responsible release, access control, and the reproducibility of findings?
    \answerYes{Yes. All data comes from publicly available Wikipedia articles. We documented our data collection and processing procedures. We released code and supporting materials to facilitate reproducibility. No personally identifiable information is collected or released.}
  \item Have you read the ethics review guidelines and ensured that your paper conforms to them?
    \answerYes{Yes}
\end{enumerate}

\item Additionally, if your study involves hypotheses testing...
\begin{enumerate}
  \item Did you clearly state the assumptions underlying all theoretical results?
    \answerYes{Yes. We state assumptions about language-country alignment and describe how we operationalize "self-aligned" versus "enemy-aligned" entities.}
  \item Have you provided justifications for all theoretical results?
    \answerYes{Yes. We ground our expectations in prior work on collective memory, ingroup bias, and framing theory.}
  \item Did you discuss competing hypotheses or theories that might challenge or complement your theoretical results?
    \answerYes{Yes. We discuss how our findings complicate simple ingroup favoritism models and note that some languages frame self-aligned entities as victims rather than victors.}
  \item Have you considered alternative mechanisms or explanations that might account for the same outcomes observed in your study?
    \answerYes{We acknowledge that differences may arise from editor composition, source availability, or cultural memory traditions, and we cannot fully disentangle these factors.}
  \item Did you address potential biases or limitations in your theoretical framework?
    \answerYes{Yes. We note that connotation frames were developed primarily from English-language sources and that the multilingual extension may not capture all language-specific connotations.}
  \item Have you related your theoretical results to the existing literature in social science?
    \answerYes{Yes. We connect our findings to work on collective memory, ingroup bias in Wikipedia, and framing theory.}
  \item Did you discuss the implications of your theoretical results for policy, practice, or further research in the social science domain?
    \answerYes{Yes. We discuss implications for understanding Wikipedia's governance model and the limits of neutrality norms, and suggest directions for future research.}
\end{enumerate}

\item Additionally, if you are including theoretical proofs...
\begin{enumerate}
  \item Did you state the full set of assumptions of all theoretical results?
    \answerNA{N/A. This paper does not include theoretical proofs.}
	\item Did you include complete proofs of all theoretical results?
    \answerNA{N/A}
\end{enumerate}

\item Additionally, if you ran machine learning experiments...
\begin{enumerate}
  \item Did you include the code, data, and instructions needed to reproduce the main experimental results (either in the supplemental material or as a URL)?
    \answerYes{Yes. Link incorporated as footnote in the Methodology section.}
  \item Did you specify all the training details (e.g., data splits, hyperparameters, how they were chosen)?
    \answerYes{Yes. The Appendix and Supplementary Materials describe the BERTopic configuration, embedding model (all-mpnet-base-v2), and note that default UMAP and HDBSCAN parameters were used. The connotation frame lexicon is pre-existing and not trained.}
     \item Did you report error bars (e.g., with respect to the random seed after running experiments multiple times)?
    \answerYes{Yes. Our statistical tests report p-values and confidence intervals where appropriate. Topic modeling results are descriptive and not evaluated with error bars.}
	\item Did you include the total amount of compute and the type of resources used (e.g., type of GPUs, internal cluster, or cloud provider)?
    \answerNA{All models were run on a local computer.}
     \item Do you justify how the proposed evaluation is sufficient and appropriate to the claims made? 
    \answerYes{Yes. We use established statistical tests (Wilcoxon signed-rank) appropriate for paired, non-normal data and apply multiple comparison corrections (Benjamini-Hochberg).}
     \item Do you discuss what is ``the cost`` of misclassification and fault (in)tolerance?
    \answerNA{N/A. This is a descriptive analysis rather than a predictive classification task.}
  
\end{enumerate}

\item Additionally, if you are using existing assets (e.g., code, data, models) or curating/releasing new assets...
\begin{enumerate}
  \item If your work uses existing assets, did you cite the creators?
    \answerYes{Yes. We cite RIVETER \cite{antoniak2023riveter}, BERTopic \cite{grootendorst2022bertopic}, Sentence-BERT \cite{reimers2019sentence}, and other tools used.}
  \item Did you mention the license of the assets?
    \answerYes{Wikipedia content is available under the CC-BY-SA and the GFDL, the MediaWiki history dumps under CC0, and code by Sap et al. (2017) under GNU GPL 3}
  \item Did you include any new assets in the supplemental material or as a URL?
    \answerYes{Yes. We provide the code and supporting reproducibility materials through an OSF repository.}
  \item Did you discuss whether and how consent was obtained from people whose data you're using/curating?
    \answerNA{N/A. Wikipedia content is publicly available under Creative Commons licensing and does not require individual consent for research use.}
  \item Did you discuss whether the data you are using/curating contains personally identifiable information or offensive content?
    \answerYes{Yes. The data consists of Wikipedia article text about historical wars. While articles may reference historical figures by name, this information is already public. War-related content may include descriptions of violence, which is inherent to the subject matter.}
\item If you are curating or releasing new datasets, did you discuss how you intend to make your datasets FAIR?
    \answerNA{N/A. We are not curating or releasing any new datasets.}
\item If you are curating or releasing new datasets, did you create a Datasheet for the Dataset? 
    \answerNA{N/A. We are not curating or releasing any new datasets.}
\end{enumerate}

\item Additionally, if you used crowdsourcing or conducted research with human subjects...
\begin{enumerate}
  \item Did you include the full text of instructions given to participants and screenshots?
    \answerYes{Appendix D describes the task instructions, including the introduction to connotation frame concepts, the qualifying quiz, and the seven annotation questions per triple.}
  \item Did you describe any potential participant risks, with mentions of Institutional Review Board (IRB) approvals?
    \answerYes{The study involved annotating short text excerpts from publicly available Wikipedia articles about historical wars. We assessed the risk to participants as minimal. The study was approved by our institution's IRB (26-04-10058).}
  \item Did you include the estimated hourly wage paid to participants and the total amount spent on participant compensation?
    \answerYes{Yes. Annotators were paid \$40 each for an estimated 70 minutes of work (approximately \$34/hour). We recruited 12 annotators across four languages, for a total of \$480.}
   \item Did you discuss how data is stored, shared, and deidentified?
   \answerYes{Annotations were collected through Prolific and do not contain personally identifiable information. Annotator IDs are pseudonymized in our data}
\end{enumerate}

\end{enumerate}

%% file: 09_Appendix.tex
\appendix
\setcounter{secnumdepth}{2}
\renewcommand{\thesection}{\Alph{section}}
\renewcommand{\thefigure}{\Alph{section}.\arabic{figure}}
\renewcommand{\thetable}{\Alph{section}.\arabic{table}}
\setcounter{figure}{0}
\setcounter{table}{0}

\section{Language Assignment for Belligerent Entities}
\label{appendix:language_assignment}
To obtain the official language(s) of each belligerent entity, we connected each entity’s English Wikipedia link to its Wikidata item identifier (QID) and inspected instance of (\texttt{P31}), which specifies the ontological class of an entity (e.g., sovereign state, military organization, or political group), to determine the type of each belligerent entity:
\begin{itemize}
    \item If an entity was categorized as a former country, sovereign state, or country, we extracted its official languages from the infobox of its English Wikipedia page.
    \item If an entity was an organization or group, we inspected country (\texttt{P17}), which indicates the sovereign state to which an entity officially belongs or is administratively associated, to determine the country affiliation of the entity. We assigned its official language(s) based on the official language(s) of the affiliated country.
    \item If an entity did not have a country affiliation, we marked its official language as unknown.
\end{itemize}

\section{Implementation Details}
\label{appendix:implementation}
In our topic modeling analysis, we employed the BERTopic framework (version 0.15.0) for topic extraction and clustering. We largely followed the framework’s default configuration for all components unless otherwise specified.

Document embeddings were generated using a pre-trained SentenceTransformer model, specifically the all-mpnet-base-v2 variant. This model was selected for its strong performance in capturing semantic similarity in sentence-level representations. As the model is pre-trained, no additional training epochs or task-specific fine-tuning were applied during the embedding process.

For dimensionality reduction, BERTopic’s default UMAP configuration was used. Topic clustering was performed using HDBSCAN, a density-based clustering algorithm that is robust to noise and well-suited for discovering topics of varying sizes. We set the minimum cluster size to 20, while all other clustering-related hyperparameters were kept at their default values.

\section{OLS Formulation}
\label{appendix:ols}

\begin{equation}
\begin{aligned}
y_{ij} =\;& \beta_0 
+ \beta_1 \cdot \text{win}_{ij}
+ \beta_2 \cdot \text{loss}_{ij} \\
&+ \gamma_1 \cdot \text{revert\_rate}_{ij}
+ \gamma_2 \cdot \text{edit\_concentration}_{ij} \\
&+ \gamma_3 \cdot \log(\text{unique\_editors}_{ij})
+ \gamma_4 \cdot \log(\text{talk\_length}_{ij}) \\
&+ \gamma_5 \cdot \log(\text{section\_count}_{ij}) \\
&+ \sum_{k=1}^{K} \lambda_k \cdot \mathbb{1}[\text{language}_{ij}=k] \\
&+ \varepsilon_{ij}
\end{aligned}
\label{eq:lmm}
\end{equation}

where \(y_{ij}\) denotes one of the nine framing outcomes for battle \(i\) in language edition \(j\), including self-, enemy-, or gap-based power, agency, and sentiment scores.

\section{Translation Human Validation Study}
\label{appendix:translation}
We provide additional details on our annotation validation study to fully describe data quality and to analyze how translation may affect connotation scores.

\subsection{Language Choice}
As the Connotation Frame lexicon is based on English, non-Latin-script languages are most susceptible to translation-induced distortion. We selected Chinese, Arabic, and Japanese as our primary validation languages, as these three represent typologically distinct language families present in our dataset and pose different challenges for NLP pipelines: Arabic's morphological complexity, Chinese's lack of inflectional morphology, and Japanese's agglutinative structure. In addition, we included English as a baseline language to establish an upper bound on annotation quality under ideal conditions, where no translation is involved and both the source text and connotation score labels are presented in the annotators' native language.

\subsection{Task Instructions}
For each language, 50 SVO triples were randomly sampled from our corpus. Annotators were recruited through Prolific\footnote{\url{https://www.prolific.com/}}. The study was hosted on our institution's Qualtrics website. We did not collect any personally identifiable information for Prolific participants. Upon entering the study, annotators first viewed an introduction page describing the task, followed by a consent form; participants who declined were redirected to Prolific. This study was approved by our institution's IRB.

Before the annotation task began, annotators received a detailed introduction to the concepts of power, agency, and sentiment as operationalized in connotation frames, accompanied by 10 illustrative examples. Annotators were then required to pass a five-question quiz with at least 80\% accuracy before proceeding. Those who failed were immediately excluded from the study. Four quiz questions were drawn directly from the provided examples, and one was closely modeled on them. The quiz was designed to ensure that annotators had sufficiently understood the instructions before beginning the task. To monitor annotation consistency, one identical triple was embedded in both the first and second halves of the task, and annotators' responses to this repeated item served as an attention check.

For each SVO triple, annotators were presented with: (1) the original sentence in the source language, along with two sentences of preceding and following context; (2) the English translation of the sentences; (3) the article title and section in the original language for broader context; and (4) the extracted Subject, Verb, and Object in English. The target sentence was typographically highlighted to facilitate identification. Annotators answered seven questions for each triple. Table~\ref{tab:annotation_questions} summarizes the seven annotation questions and their corresponding response options. The English condition was an exception: Q3 (Verb Translation Quality) was omitted because the articles were originally written in English and no translation was involved.

\begin{table*}[h]
\centering
\footnotesize
\setlength{\tabcolsep}{5pt}
\renewcommand{\arraystretch}{1.2}

\begin{tabular}{p{0.05\textwidth} p{0.30\textwidth} p{0.59\textwidth}}
\toprule
\textbf{Question} & \textbf{} & \textbf{Response Options} \\
\midrule

Q1 &
\textbf{Subject Identification}

Is the Subject correctly identified?
&
Yes 

Incorrect or partially correct. Please type the correct Subject. \\
\midrule

Q2 &
\textbf{Object Identification}

Is the Object correctly identified?
&
Yes 

Incorrect or partially correct. Please type the correct Object. \\
\midrule

Q3 &
\textbf{Verb Translation Quality}

How accurately does the translated verb preserve the meaning of the source verb?
&
The translated verb fully preserves the meaning and connotative direction of the source verb

Minor deviation: The translated verb is semantically close but has a slightly different connotation or intensity

Moderately accurately

Major deviation: The translated verb has a noticeably different meaning that could alter the connotation scores

Wrong/Missing: The verb is completely mistranslated, untranslated, or absent \\
\midrule

Q4 &
\textbf{Power}

Does the Subject have more or less power than the Object?
&
Subject has more power than Object

Subject and object have equal power

Subject has less power than Object \\
\midrule

Q5 &
\textbf{Agency}

Does the verb suggest that the Subject is acting with intention and control?
&
High agency — the Subject acts deliberately

Neutral

Low agency — the Subject is passive or acted upon \\
\midrule

Q6 &
\textbf{Sentiment toward Subject}

How does the writer seem to view the Subject?
&
Positive

Neutral

Negative \\
\midrule

Q7 &
\textbf{Sentiment toward Object}

How does the writer seem to view the Object?
&
Positive

Neutral

Negative \\
\bottomrule
\end{tabular}
\caption{Annotation questions and response options used in the translation human validation study.}
\label{tab:annotation_questions}
\end{table*}

Prior to releasing tasks on Prolific, a native speaker proficient in English verified that each original sentence and its translation were correctly aligned, and that the highlighted sentence corresponded to the correct triple. This verification step was not applied to the English condition, where the source and presented text are identical, and no translation alignment needed to be confirmed.

\subsection{Task Settings}
To ensure annotation quality, we imposed geographic restrictions on the annotator pool: Japanese annotators were restricted to residents of Japan, and Chinese annotators to residents of China. English annotators were not subject to geographic restrictions, but they were required to be native English speakers. All annotators were required to be proficient in English, as the extracted SVO triples and connotation score labels were presented in English. Payment was set at \$40 per annotator, based on an estimated completion time of approximately 70 minutes determined during pilot testing with three native speakers who were proficient in English. We recruited 3 annotators per language, for a total of 12 across four languages. All annotators passed the initial quiz and the in-task attention check.

\subsection{Annotator Quality Control}
Following \citet{park2021multilingual}, we assessed annotator reliability by computing each annotator's disagreement rate: the proportion of instances on which that annotator's response differed from both other annotators. For Chinese, Arabic, and Japanese, disagreement rates were computed over all 50 items $\times$ 7 questions = 350 judgments per annotator. For English, disagreement rates were computed over 50 items $\times$ 6 questions = 300 judgments per annotator. We compared each annotator's rate against the mean and standard deviation within their language group, retaining annotators within one standard deviation of the mean. One Japanese annotator exceeded this threshold and was excluded from the final analysis; all remaining annotators were retained.

\subsection{Inter-Annotator Agreement}

Table~\ref{tab:iaa} reports Krippendorff's $\alpha$ for each question and language. We use nominal $\alpha$ for Q1 and Q2 (binary categorical) and ordinal $\alpha$ for Q3--Q7 (ordered scales). For Q1 and Q2, where percent agreement was generally high, low $\alpha$ values reflect the well-known restricted-range artifact: when the distribution is heavily skewed toward one response, expected chance agreement approaches observed agreement, depressing $\alpha$ regardless of true annotator reliability \citep{krippendorff2011computing}. We therefore report percent agreement alongside $\alpha$.

\begin{table}[h]
\centering
\footnotesize
\begin{tabular}{llrr}
\toprule
\textbf{Language} & \textbf{Question} & {$\alpha$} & {\%Agr} \\
\midrule
\multirow{7}{*}{Arabic}
 & Q1 Subject ID    & .178    & 92\% \\
 & Q2 Object ID     & .324    & 96\% \\
 & Q3 Verb Quality  & .489    & 90\% \\
 & Q4 Power         & .514    & 76\% \\
 & Q5 Agency        & .425    & 82\% \\
 & Q6 Sent.\ (Subj) & .441    & 34\% \\
 & Q7 Sent.\ (Obj)  & .151    & 58\% \\
\midrule
\multirow{7}{*}{Chinese}
 & Q1 Subject ID    & .483    & 94\% \\
 & Q2 Object ID     & .521    & 84\% \\
 & Q3 Verb Quality  & .049    & 48\% \\
 & Q4 Power         & .194    & 74\% \\
 & Q5 Agency        & .309    & 74\% \\
 & Q6 Sent.\ (Subj) & .239    & 22\% \\
 & Q7 Sent.\ (Obj)  & .216    & 58\% \\
\midrule
\multirow{7}{*}{Japanese}
 & Q1 Subject ID    & $-$.020 & 96\% \\
 & Q2 Object ID     & $-$.075 & 86\% \\
 & Q3 Verb Quality  & .171    & 68\% \\
 & Q4 Power         & .306    & 74\% \\
 & Q5 Agency        & .206    & 75\% \\
 & Q6 Sent.\ (Subj) & $-$.179 & 33\% \\
 & Q7 Sent.\ (Obj)  & $-$.132 & 60\% \\
\midrule
\multirow{7}{*}{English}
 & Q1 Subject ID    & .625    & 92\% \\
 & Q2 Object ID     & .464    & 82\% \\
 & Q3 Verb Quality  & —       & —    \\
 & Q4 Power         & .168    & 72\% \\
 & Q5 Agency        & .224    & 70\% \\
 & Q6 Sent.\ (Subj) & .220    & 58\% \\
 & Q7 Sent.\ (Obj)  & .050    & 46\% \\
\bottomrule
\end{tabular}
\caption{Inter-annotator agreement per question and language.
$\alpha$: Krippendorff's alpha (nominal for Q1--Q2, ordinal for Q3--Q7).
\%Agr: percentage of items with full three-annotator agreement.
For Japanese, values reflect the two retained annotators after one was excluded.
Q3 is not applicable (—) for English as no translation is involved.
Low $\alpha$ on Q1--Q2 reflects a restricted-range artifact given near-ceiling accuracy rates.}
\label{tab:iaa}
\end{table}

\subsection{Error Propagation Analysis}

To examine whether verb translation quality moderates the correspondence between human-assigned connotation labels and RIVETER lexicon scores, we stratified items by translation quality (Perfect, Minor, Major) and SVO accuracy (Accurate, Inaccurate), treating subject and object separately. As no triples received a Wrong rating in any of the three languages, this category was excluded from the analysis. Following \citet{park2021multilingual}, we mapped each annotator's judgment to a $(-1, 0, +1)$ value and averaged scores across annotators, then ternarized the aggregated score using the boundaries positive $[0.35, 1]$, neutral $(-0.35, 0.35)$, and negative $[-1, -0.35]$. Table~\ref{tab:error_propagation_combined} shows the consistency between human annotations and RIVETER lexicon scores across Arabic, Chinese and Japanese. Note that the RIVETER lexicon maintains separate coverage for power/agency and sentiment dimensions; not all verbs are assigned sentiment scores, which accounts for the reduced sample sizes observed in sentiment consistency cells relative to power and agency.

For Power and Agency dimensions, Arabic and Chinese show consistently higher agreement with lexicon scores than Japanese across all conditions, with Perfect-translation subject-accurate triples reaching 78.3\%/80.4\%, 80.0\%/80.0\%, and 70.0\%/65.4\% on Power/Agency respectively. Crucially, introducing Minor translation deviation does not substantially alter agreement relative to Perfect translation across any language. For the Chinese and Japanese triples where annotators disagreed on translation quality, all had accurate SVO extraction, and each triple received one vote for each of the three quality levels (Perfect, Minor, and Major). Agreement rates on Power and Agency are broadly comparable to those observed in the majority-vote Perfect and Minor groups, suggesting that annotators' assessments of power and agency implications remain relatively consistent even when they disagree on translation fidelity. Across all four languages, chi-square tests revealed no significant cross-linguistic differences in overall Power or Agency consistency (all $\chi^2(3)$ ns after Bonferroni correction), suggesting that these two dimensions are captured comparably well by the RIVETER lexicon regardless of language or translation quality.

For Sentiment, consistency rates are substantially lower than Power and Agency across all conditions and languages. This is consistent with the low inter-annotator agreement on sentiment dimensions throughout our validation, and likely reflects both the subjective nature of sentiment attribution and the cultural specificity of affective presuppositions in war narratives that the English-based RIVETER lexicon does not capture. Cross-linguistic differences in Sentiment were significant only for Object Sentiment ($\chi^2(3) = 18.62$, $p < .001$), with Arabic showing significantly higher consistency than English after Bonferroni correction ($\text{OR} = 6.06$, $p_{\text{adj}} = .001$). No significant cross-linguistic differences were observed for Subject Sentiment. Sentiment-based findings should therefore be interpreted with substantial caution.

\begin{table*}[h]
\centering
\scriptsize
\setlength{\tabcolsep}{4pt}
\renewcommand{\arraystretch}{0.9}
\begin{tabular}{l||cccc||cccc||cccc}
\toprule
& \multicolumn{4}{c||}{\textbf{Arabic}} & \multicolumn{4}{c||}{\textbf{Chinese}} & \multicolumn{4}{c}{\textbf{Japanese}} \\
\textbf{Group} & Pow. & Agen. & Sent$_\text{s}$ & Sent$_\text{o}$ & Pow. & Agen. & Sent$_\text{s}$ & Sent$_\text{o}$ & Pow. & Agen. & Sent$_\text{s}$ & Sent$_\text{o}$ \\
\hline
Perfect $|$ Subj.\ Acc.   & 78.3(46) & 80.4(46) & 35.1(38) & —        & 80.0(40) & 80.0(40) & 41.7(36) & —        & 70.0(26) & 65.4(26) & 61.5(26) & —        \\
Perfect $|$ Subj.\ Inacc. & —        & —        & —        & —        &  0.0(1)  &  0.0(1)  &  0.0(1)  & —        & —        & —        & —        & —        \\
Perfect $|$ Obj.\ Acc.    & 76.6(46) & 78.2(46) & —        & 78.9(38) & 80.6(36) & 77.8(36) & —        & 61.3(31) & 73.9(25) & 67.8(25) & —        & 62.5(23) \\
Perfect $|$ Obj.\ Inacc.  & —        & —        & —        & —        &  0.0(5)  &  0.0(5)  & —        &  0.0(2)  &  0.0(1)  &  0.0(1)  & —        &  0.0(1)  \\
\hline\hline
Minor $|$ Subj.\ Acc.     & 100.0(2) & 50.0(2)  & 50.0(2)  & —        & 50.0(2)  & 50.0(2)  & 50.0(2)  & —        & 75.0(4)  & 75.0(4)  & 50.0(4)  & —        \\
Minor $|$ Subj.\ Inacc.   & —        & —        & —        & —        & —        & —        & —        & —        & —        & —        & —        & —        \\
Minor $|$ Obj.\ Acc.      & 100.0(2) & 50.0(2)  & —        & 50.0(2)  & 100.0(2) & 100.0(2) & —        &  0.0(2)  & 33.3(3)  & 33.3(3)  & —        & 33.3(3)  \\
Minor $|$ Obj.\ Inacc.    & —        & —        & —        & —        & —        & —        & —        & —        & —        & —        & —        & —        \\
\hline\hline
Major $|$ Subj.\ Acc.     & —        & —        & —        & —        & —        & —        & —        & —        & 100.0(1) & 100.0(1) &  0.0(1)  & —        \\
Major $|$ Subj.\ Inacc.   &  0.0(1)  &  0.0(1)  & —        & —        & —        & —        & —        & —        & —        & —        & —        & —        \\
Major $|$ Obj.\ Acc.      & —        & —        & —        & —        & —        & —        & —        & —        & 100.0(1) & 100.0(1) & —        & 100.0(1) \\
Major $|$ Obj.\ Inacc.    &  0.0(1)  &  0.0(1)  & —        & —        & —        & —        & —        & —        & —        & —        & —        & —        \\
\bottomrule
\end{tabular}
\caption{Consistency (\%) between human annotations and RIVETER lexicon scores, stratified by translation quality, SVO axis, and accuracy condition. Values are formatted as \textit{accuracy\%}(\textit{n}). Sent$_\text{s}$: sentiment toward subject; Sent$_\text{o}$: sentiment toward object. Cells marked with --- indicate zero instances or non-applicable dimensions.}
\label{tab:error_propagation_combined}
\end{table*}

\begin{table}[h]
\centering
\small
\begin{tabular}{lrrrr}
\toprule
\textbf{Group} & \textbf{Pow.} & \textbf{Agen.} & \textbf{Sent\textsubscript{s}} & \textbf{Sent\textsubscript{o}} \\
\midrule
Subj.\ Accurate   & 73.9(46) & 69.6(46) & 31.0(38) & 42.9(38) \\
Subj.\ Inaccurate &  0.0(4)  &  0.0(4)  &  0.0(4)  &  0.0(4)  \\
Obj.\ Accurate    & 72.3(47) & 68.1(47) & 32.4(47) & 35.3(37) \\
Obj.\ Inaccurate  &  0.0(3)  &  0.0(3)  &  0.0(3)  &  0.0(3)  \\
\bottomrule
\end{tabular}
\caption{Consistency (\%) between human annotations and RIVETER lexicon scores for English, stratified by SVO accuracy condition. Sent\textsubscript{s}: sentiment toward subject; Sent\textsubscript{o}: sentiment toward object.}
\label{tab:consistency_english}
\end{table}

\begin{table}[h]
\centering
\small
\begin{tabular}{lcccc}
\toprule
& \textbf{Pow.} & \textbf{Agen.} & \textbf{Sent$_\text{s}$} & \textbf{Sent$_\text{o}$} \\
\midrule
Chinese ($n$=7)   & 71.4 & 42.8 & 14.3 & 28.6 \\
Japanese ($n$=16) & 68.7 & 62.5 & 45.5 & 45.7 \\
\bottomrule
\end{tabular}
\caption{Consistency (\%) for Q3 cases with no majority agreement (Chinese: all three annotators disagreed; Japanese: both annotators disagreed). Arabic is excluded as it had no such triples.}
\label{tab:verb_disagreement_consistency}
\end{table}

\begin{table}[h]
\centering
\small
\begin{tabular}{lrrrr}
\toprule
\textbf{Dimension} & \textbf{Arabic} & \textbf{Chinese} & \textbf{Japanese} & \textbf{English} \\
\midrule
Subject / Power           & 77.6 & 76.0 & 73.9 & 68.0 \\
Subject / Agency          & 75.5 & 74.0 & 69.0 & 64.0 \\
Subject / Sent\textsubscript{s} & 35.8 & 37.0 & 53.8 & 28.9 \\
\midrule
Object / Power            & 76.1 & 72.1 & 69.0 & 68.0 \\
Object / Agency           & 74.0 & 66.0 & 70.1 & 64.0 \\
Object / Sent\textsubscript{o}  & 77.5 & 50.0 & 55.1 & 36.6 \\
\bottomrule
\end{tabular}
\caption{Overall consistency (\%) across four languages, stratified by SVO axis and dimension. Sent\textsubscript{s}: sentiment toward subject; Sent\textsubscript{o}: sentiment toward object. For Japanese, results reflect the two retained annotators after one was excluded.}
\label{tab:consistency_overall}
\end{table}

\subsection{Robustness Check: Alternative Translation Service}

To assess the robustness of our findings to the choice of translation pipeline, we replicated the RQ1 analysis using the Wikimedia Foundation's MinT translation service\footnote{\url{https://www.mediawiki.org/wiki/MinT}} as an alternative to the Google Translate API. MinT is an open-source, Wikipedia-native machine translation system, and its use as an independent robustness check helps rule out the possibility that our results are artifacts of a single translation engine. The replication yielded results consistent with the main analysis, as shown in Tables~\ref{tab:self_greater_main_mint} and~\ref{tab:self_less_main_mint}, supporting the validity of our findings across translation pipelines.

\begin{table}[h]
\centering
\small
\begin{tabular}{llrr}
\toprule
\textbf{Language} & \textbf{Dimension} & \textbf{$p$} & \textbf{Sig.} \\
\midrule
English  & Agency    & $<0.01$   & **  \\
\midrule
\multirow{2}{*}{Hebrew}
 & Power   & 0.002     & **  \\
 & Agency  & $<0.001$  & *** \\
\midrule
\multirow{2}{*}{Japanese}
 & Power   & $<0.001$  & *** \\
 & Agency  & $<0.001$  & *** \\
\midrule
\multirow{3}{*}{Russian}
 & Power     & 0.016    & *   \\
 & Agency    & $<0.001$ & *** \\
 & Sentiment & 0.034    & *   \\
\bottomrule
\end{tabular}
\caption{Wilcoxon signed-rank test results (MinT robustness check) for languages in which self-aligned entities are portrayed as \textit{more} powerful, \textit{more} agentic, or \textit{more positive} than enemy-aligned entities. Benjamini--Hochberg false discovery rate correction was applied. $p$-values are two-sided. Significance levels: *** $p<0.001$, ** $p<0.01$, * $p<0.05$.}
\label{tab:self_greater_main_mint}
\end{table}

\begin{table}[h]
\centering
\small
\begin{tabular}{llrr}
\toprule
\textbf{Language} & \textbf{Dimension} & \textbf{$p$} & \textbf{Sig.} \\
\midrule
\multirow{2}{*}{Arabic}
 & Power   & $<0.001$ & *** \\
 & Agency  & $<0.001$ & *** \\
\midrule
\multirow{2}{*}{Ukrainian}
 & Power   & 0.008    & **  \\
 & Agency  & 0.015    & **  \\
\midrule
\multirow{2}{*}{Vietnamese}
 & Power   & $<0.001$ & *** \\
 & Agency  & $<0.001$ & *** \\
\bottomrule
\end{tabular}
\caption{Wilcoxon signed-rank test results (MinT robustness check) for languages in which self-aligned entities are portrayed as \textit{less} powerful, \textit{less} agentic, or \textit{more negative} than enemy-aligned entities. Benjamini--Hochberg false discovery rate correction was applied. $p$-values are two-sided. Significance levels: *** $p<0.001$, ** $p<0.01$, * $p<0.05$.}
\label{tab:self_less_main_mint}
\end{table}

\section{Mixed-Effects Regression Robustness Check}
\label{appendix:mixed-effects}
As a robustness check, we re-estimated the regression models using linear mixed-effects models with language specified as a random intercept to account for potential clustering within language editions. The dependent variables remained the same as in the main OLS analysis, including self-, enemy-, and gap-based framing scores across the power, agency, and sentiment dimensions. Fixed effects included battle outcomes (win and loss, with draw as the reference category) and the same editorial behavior variables used in the main models.

The mixed-effects specification is given by:

\[
y_{ij}
=
\beta_0
+
\beta_1 \text{Win}_{ij}
+
\beta_2 \text{Loss}_{ij}
+
\sum_k \beta_k X_{kij}
+
u_j
+
\epsilon_{ij}
\]

where \(y_{ij}\) denotes the framing outcome for battle \(i\) in language edition \(j\), \(X_{kij}\) represents the editorial behavior covariates, \(u_j\) is a language-specific random intercept, and \(\epsilon_{ij}\) is the residual error term.

Continuous predictors were standardized prior to estimation, and count-based variables were log-transformed to reduce skewness. We further examined model assumptions by inspecting residual normality, skewness, kurtosis, Q--Q plots, heteroscedasticity, and multicollinearity diagnostics.

Overall, the mixed-effects models produced substantively similar results to the OLS models reported in the main text, suggesting that the observed framing patterns are robust to alternative model specifications that account for language-level clustering.

\begin{table*}[!t]
\centering
\scriptsize
\setlength{\tabcolsep}{4pt}
\renewcommand{\arraystretch}{1.1} 

\caption{
Linear mixed-effects regression results for self-, enemy-, and gap-framing scores across power, agency, and sentiment dimensions. Language was modeled as a random intercept. Continuous predictors were standardized prior to estimation.
}
\label{tab:mixed_effects_models}

\begin{tabular}{lccccccccc}
\hline

& \multicolumn{3}{c}{\textbf{Power}}
& \multicolumn{3}{c}{\textbf{Agency}}
& \multicolumn{3}{c}{\textbf{Sentiment}} \\

\cline{2-10}

\textbf{Predictor}
& \textbf{Self}
& \textbf{Enemy}
& \textbf{Gap}
& \textbf{Self}
& \textbf{Enemy}
& \textbf{Gap}
& \textbf{Self}
& \textbf{Enemy}
& \textbf{Gap} \\

\hline

\multicolumn{10}{l}{\textit{Battle Outcomes \& Editorial Behaviors}} \\
\hline

Win
& ---
& -0.096***
& +0.119**
& ---
& -0.082***
& +0.079**
& ---
& -0.078***
& +0.102*** \\

Loss
& -0.090***
& +0.070*
& -0.161***
& -0.095***
& ---
& -0.123***
& ---
& -0.032.
& --- \\

Revert rate (z)
& ---
& ---
& ---
& ---
& -0.012.
& ---
& ---
& ---
& --- \\

Edit concentration (z)
& ---
& ---
& ---
& ---
& -0.026***
& +0.021.
& ---
& -0.017*
& +0.016. \\

Log unique editors (z)
& +0.036**
& ---
& ---
& ---
& ---
& ---
& ---
& +0.017.
& --- \\

Section log (z)
& ---
& ---
& ---
& -0.022**
& ---
& -0.022.
& +0.026**
& +0.025**
& --- \\

Talk log (z)
& ---
& ---
& ---
& +0.014.
& ---
& +0.020.
& ---
& ---
& +0.020. \\

\hline
\multicolumn{10}{l}{\textit{Random Effects \& Model Diagnostics}} \\
\hline

Language RE Variance ($\sigma_u^2$)
& 0.0113 & 0.0030 & 0.0308
& 0.0065 & 0.0018 & 0.0130
& 0.0003 & 0.0015 & 0.0019 \\

Language RE SD ($\sigma_u$)
& 0.1061 & 0.0546 & 0.1755
& 0.0804 & 0.0423 & 0.1139
& 0.0176 & 0.0384 & 0.0437 \\

Residual Variance ($\sigma_\epsilon^2$)
& 0.2447 & 0.2930 & 0.6096
& 0.1346 & 0.1463 & 0.2883
& 0.1334 & 0.1448 & 0.2376 \\

Residual SD ($\sigma_\epsilon$)
& 0.4946 & 0.5413 & 0.7808
& 0.3669 & 0.3825 & 0.5369
& 0.3652 & 0.3805 & 0.4874 \\

ICC
& 4.40\% & 1.01\% & 4.81\%
& 4.58\% & 1.21\% & 4.31\%
& 0.23\% & 1.01\% & 0.80\% \\

\hline

\multicolumn{10}{l}{\scriptsize Significance: *** $p<.001$, ** $p<.01$, * $p<.05$, . $p<.10$} \\

\hline
\end{tabular}
\end{table*}